\documentclass[aps,prx,twocolumn,floats,showpacs,superscriptaddress,nofootinbib,longbibliography]{revtex4-2}
\usepackage{graphicx}
\usepackage{times,bbm}
\usepackage{graphics,dcolumn,bm,float}
\usepackage{amssymb,amsmath,rotate,color,amsfonts}
\usepackage[title,titletoc,toc]{appendix}
\usepackage{mathtools}
\usepackage{booktabs}
\usepackage{tcolorbox}
\usepackage[pagebackref=false,colorlinks,linkcolor=magenta,citecolor=blue,urlcolor=magenta]{hyperref}

\usepackage{wrapfig}
\usepackage{lipsum}
\usepackage{mwe}
\usepackage[mathlines]{lineno}
\usepackage[mathscr]{euscript}
\usepackage{hyperref}
\usepackage{breqn}
\usepackage{bbold}
\usepackage{pgfplots}
\usepackage{float}
\usepackage{tkz-euclide}
\usepackage{braket}
\usepackage{physics}
\usepackage[caption=false]{subfig}
\usepackage[export]{adjustbox}
\usepackage{tikz}
\usepackage{slashed}
\usepackage{bm}
\usetikzlibrary{through,calc}
\usetikzlibrary{positioning}

\newcommand{\be}{\begin{equation}}
\newcommand{\ee}{\end{equation}}

\newcommand{\bea}{\begin{eqnarray}}
\newcommand{\eea}{\end{eqnarray}}

\newcommand{\bmt}{\left[\begin{matrix}}
\newcommand{\emt}{\end{matrix}\right]}

\begin{document}
\preprint{}
\title{
Solid-State Gravitational Redshift: Transport Signatures of Massless Dirac Fermions in Tilted Dirac Cone Heterostructures}
\author{Marziyeh Karmand} \affiliation{ Faculty of Physics$,$ University of Isfahan$,$ Isfahan 81746-73441$,$ Iran}
\author{Mohsen Amini}\email{msn.amini@sci.ui.ac.ir} \affiliation{ Faculty of Physics$,$ University of Isfahan$,$ Isfahan 81746-73441$,$ Iran}
\author{Morteza Soltani} \affiliation{ Faculty of Physics$,$ University of Isfahan$,$ Isfahan 81746-73441$,$ Iran}
\author{Ebrahim Ghanbari-Adivi}\affiliation{ Faculty of Physics$,$ University of Isfahan$,$ Isfahan 81746-73441$,$ Iran}
\author{Seyed Akbar Jafari}\affiliation{ II. Physikalisches Institute$,$ RWTH Aachen University$,$ 52074 Aachen$,$ Germany}
\date{\today}

\begin{abstract}
We investigate the propagation of electron waves in a two-dimensional tilted Dirac cone heterostructure where tilt depends on the coordinate $z$ along 
the junction. The resulting Dirac equation in an emergent curved spacetime for the spinor $\psi(z)$ can be efficiently solved using 4th-order Runge-Kutta 
numerical method by a transformation to a "suitable" spinor $\varphi$ where the resulting Dirac cone looks locally upright. 
The spatial texture of the tilt induces oscillatory behaviors in key physical quantities such as the norm \(|\varphi(z)|^2\) of the wave function, 
polar and azimuthal angles \(\Theta(z)\) and \(\Phi(z)\) of the pseudospin, and the integrated transmission \(\tau\) where oscillation 
wave-lengths get shorter (longer) in stronger (weaker) tilt regions. Such an oscillatory behavior reminiscent of {\em gravitational red-shift} 
is an indicator of an underlying spacetime metric that can be probed in tunneling experiments. 
We derive analytical approximations for the position-dependent wave numbers \(\Delta k_z(z)\) that explain the red-shift patterns 
and corroborate them with numerical simulations. 
For a tilt "bump" spread over length scale $\ell$, upon increasing $\ell$, the {\em amplitude} of red-shifted oscillations reduces whereas the number of peaks increases. 
The scale invariance of Dirac equation allows to probe these aspects of $\ell$-dependence by a voltage sweep in transmission experiments. 
Smooth variations of the tilt reduce impedance mismatch of the electron waves, thereby giving rise to very high transmission rate.
This concept can be used in combination with a sigmoid-shaped tilt texture for red-shift or blue-shift engineering of the transmitted
waves, depending on whether the sigmoid is downswing or upswing. 
\end{abstract}



\maketitle
\narrowtext

\section{Introduction~\label{Sec01}}
Tilted Dirac fermions~\cite{Muechler2016} can arise when two bands arbitrarily cross. They also appear in many two-dimensional system~\cite{Isobe2017} from organic conductors~\cite{Kobayashi2007} to $(110)$ surface of elemental tungsten~\cite{Varykhalov2017} to $8pmmn$ borohpene~~\cite{LopezBezanilla2016}. 

Tilting as a band structure aspect can drive many interesting properties such as tunneling valley Hall effect~\cite{Zhang2023}, anisotropic optical conductivity~\cite{Tan2021}, plasmonic gains~\cite{Park2022}, tilt-induced kind in the plasmon modes~\cite{JalaliMola2018,JalaliMola2018double}, hyperbolic plasmon modes~\cite{Torbatian2021,Mojarro2022}, anomalous heat flow~\cite{Sengupta2018}. While the Dirac point in the non-interacting systems is protected by appropriately generalized chiral symmetry~\cite{KAWARABAYASHI2012}, upon including many-body interactions, interesting phenomena such as chiral excitionic instability~\cite{Ohki2020}, chiral symmetry breaking~\cite{Gomes2021}, 
maybe possible. Also adding disorder and interactions sets interesting competition between the tilt parameter and Coulomb interactions~\cite{Zhao2019}. 
The tilt effects shows up in many measurements such as thermal difference reflectivity~\cite{Mojarro2023}, RKKY interaction mediated by tilted Dirac fermions~\cite{Paul2019},
spin transport~\cite{Sinha2019}, in-plane conductivity~\cite{Suzumura2014}, 

While upright Dirac cones represent solid-state realization of an emergent Lorentz invariance which is associated with the corresponding
emergent Minkowski spacetime with coordinates $x^\mu=(v_Ft,x^1,\ldots,x^d)$\footnote{Here $d$ is the spatial dimension.} and the metric $ds^2=-v_F^2dt^2+d{\vec x}^2$, the tilted Dirac fermions are associated with the metric $ds^2=-v_F^2dt^2+(d{\vec x}-{\vec\zeta}v_Fdt)^2$~\cite{Volovik2017} known as Painlev\'e-Gulstrand spacetime. In this way interesting 
aspects of curved spacetime are brought into solids possessing a tilting texture ${\vec\zeta}({\vec x})$~\cite{Volovik2021,Liang2019,Zubkov2018,Volovik2016,Kedem2020,Konye2022,Farajollahpour2020,Hashimoto2020}.
The structure of the above Painlev\'e-Gulstrand (PG) spacetime is defined by shape of the function $\zeta$ in space or time. $|\zeta|<1$ defines exterior of blackhole that upon 
crossing the event horizon at $\zeta=1$ enters the interior of blackhole with $\zeta>1$~\cite{Zubkov2018,Volovik2016,Kedem2020,Konye2022}. 

The tilt of the Dirac cone can be manipulated which will then correspond solid-state physics way of controling the above PG metric. 
For example in $8Pmmn$ borophene it has been shown that replacing boron atoms arranged on chain-like threads e.g. by carbon atoms significantly 
modifies the tilting~\cite{Yekta2023}. As such a texture of dopants (random or deterministic) will generate a corresponding curved 
spacetime that can be random~\cite{Ghorashi2020} or deterministic.
Other class of Dirac cones that can be manipulated by external magnetic influence are spin-orbit coupled Dirac cones e.g. on the surface of topological 
insulators~\cite{Ogawa2016}. In such systems a textured magnetic influence such as a domain wall deposited on the surface of a topological insulator or the 
$(110)$ surface of tungsten~\cite{Varykhalov2017} are expected to create spatial variations by tens of percents of tilt~\cite{Jafari2024}. 

\begin{figure}[t!]  
\begin{center}  
\includegraphics[width=0.50\textwidth]{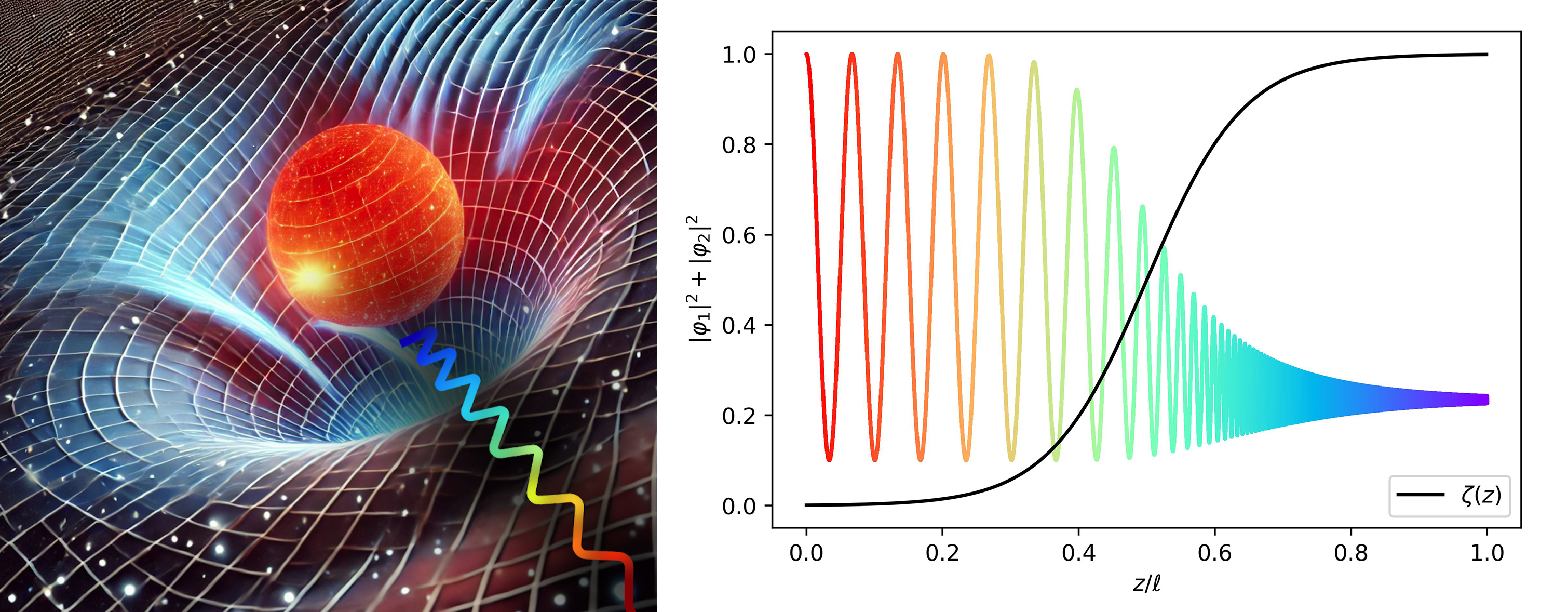}  
\end{center}  
\caption{(left) Schematic representation of the gravitational red-shift: 
Although photons do not have mass, as a result of curved spacetime, emitted photon near a gravitational source looses its "kinetic energy" 
$h\nu$ as it overcomes the gravity to move away from stronger gravity regions to weaker gravity regions, thereby looking red in less gravitating region. 
(right) Probability distribution for an electron crossing a region with a tilting profile varying from $\zeta=0$ to $\zeta=0.999$. In smaller 
$\zeta$ region the oscillations have longer wave length (i.e. red-shifted) which can be detected in conductance. 
}  
\label{FigRedshift}
\end{figure}

Given the above possibility of synthesizing PG metric in solids, how do the corresponding "gravity-like" aspects manifest in spectroscopies?
One of the salient predictions of
Einstein's general relativity was that the photons (despite being massless), when climb up a gravitational potential, loose their energy $h\nu$, thereby 
getting red-shifted~\cite{Einstein1916} as in Fig.~\ref{FigRedshift}. Within Newtonian mechanics this would make sense only if the photon had mass. 
For a massless particle, the gravitational red-shift can be explained in terms of an underlying curved spacetime. 
This tiny shift~\cite{Ryder2009} involving fractional frequency changes of the order $10^{-15}$ was measured in 1959 by Pound and Rebka~\cite{Pound1959} 
using high-accuracy measurement enabled by M\"ossbauer effect. 

The purpose of this paper is to theoretically demonstrate a similar effect in a solid-state setup for massless Dirac electrons 
{\em without using general relativity's computational tools}. The remarkable aspect of the present solid-state gravitational red-shift
is that the fractional change in this case can be on the scale of $\sim 1$ or larger. 
Such a red-shift like behavior of Dirac electrons in solids can be similarly considered to be one possible indicator
of an underlying emergent spacetime geometry. 
A change of variable from the original Dirac spinor $\psi(z)$ to a convenient spinor $\varphi(z)$ acts like 
a change of frame of reference from "substrate frame" where the transport experiment is being conducted, to a locally non-tilted frame of reference. 

Recent advance in the theoretical interpretation of zero-bias photocurrents in spin-orbit-coupled surface Dirac cones of topological 
insulators -— attributed to frame transformations~\cite{Jafari2024}, promises exciting prospects for inducing strong position-dependent tilts via magnetic domain walls.
The smooth variation of tilt not only minimizes impedance mismatch, leading to nearly perfect transmission, but also unlocks the potential for tilted Dirac cone junctions 
to act as tunable devices—capable of red- or blue-shifting electron waves depending on whether the tilt profile follows an upswing or downswing along the junction.
This introduces a novel concept for wavelength up- or down-conversion in {\em conducting} media, harnessing abstract gravitational concepts to enable useful 
applications -- distinct from traditional semiconductor-based approaches.

The paper is organized as follows: 
In section~\ref{Sec02} we discuss the boundary conditions and the transformation that simplifies the Dirac equation in variable tilt background. 
In section~\ref{Sec03} we formulate the transport coefficients in the transformed spinor basis. 
Section~\ref{Sec04} we present numerical results and our WKB-like approximate analytic treatment. 
In section~\ref{Sec05} we compute the transmission coefficient that is accessible in transport measurements. 
We end the paper in section~\ref{Sec06} by a summary and outlook. 

\section{Alternative representation of the Hamiltonian\label{Sec02}}

In this section, we derive an alternative representation of the Hamiltonian that can comfortably give the continuity equation necessary to describe the behavior of massless 
Dirac fermions in two-dimensional (2D) materials with position-dependent tilting. Let us start with the conventional expression of the Hamiltonian for massless Dirac 
fermions in a 2D system with a constant tilt the \( z \)-direction~\cite{Jafari2024}  
\begin{equation}\label{Eq01}\begin{split}
{\cal H} & = {v}_F \sigma_x k_x + \Big({v}_F \sigma_z + {v}_t \sigma_0\Big)k_z  \\ &= - i {v}_F \sigma_x {\partial\over
\partial x} - i \Big({v}_F \sigma_z + {v}_t \sigma_0\Big){\partial\over
\partial z},
\end{split}
\end{equation}
where \({v}_F\) and \({v}_t\) denote the Fermi and tilt velocities, respectively, while \(\sigma_x\) and \(\sigma_z\) represent the well-known Pauli matrices, 
and \({\sigma_0}\) is the \(2 \times 2\) identity matrix. For later convenience we have rotated the standard representation of the Dirac equation in the
$xy$ plan around the $x$ axis to obtain a representation in the $xz$ plane. This form of the Hamiltonian has the advantage that it can be rewritten as 
\begin{equation}\label{Eq02}
{\cal H} = - i {v}_F \Big(\sigma_x {\partial\over \partial x} + {\cal M} \sigma_z {\partial\over
\partial z} {\cal M}\Big),
\end{equation}  
where  
\begin{equation}\label{Eq03}
{\cal M} = \Big[ \begin{array}{cc} \sqrt{1+\zeta} & 0 \\ 0 & \sqrt{1-\zeta} \end{array} \Big],
\end{equation}  
and \(\zeta = {v}_t / {v}_F\) is the tilting parameter that is driven by the coefficient $v_t$ of the unit matrix $\sigma_0$ that has been now
absorbed into the above matrix $\cal M$. This transformation is an essential step of the present work that facilitates derivation of analytical 
expressions, along with plausible numerical solutions of the ensuing wave equations. The above form can now be nicely generalized to 
a situation with spatially variable $\zeta$. 
As particles propagate through the medium, the Hamiltonian must be constructed to ensure the corresponding continuity equation is satisfied. With this crucial requirement in mind, and based on the Hamiltonian provided in Eq.~\eqref{Eq02}, it is reasonable to introduce the desired Hamiltonian as  
\begin{equation}
{\cal H} = - i {v}_F \Big[ \sigma_x {\partial\over \partial x} + {\cal M}(z) \sigma_z {\partial\over
\partial z} {\cal M}(z) \Big].
\label{postulate.eqn}
\end{equation}  
\par
It is important to note that, since the tilting parameter \(\zeta\) is a function of \(z\), the matrix \({\cal M}\) also depends on \(z\), and the derivative with respect 
to \(z\) acts on this matrix as well. 
This equation is similar to the way position-dependent mass in semiconducting heterostructure are handled~\cite{Koc2003,Mazhari2013,RosasOrtiz2020},
although generalizations can be possible by allowing the arbitrary powers $(1\pm\zeta)^\alpha$ with $\alpha=a,b$ for the two $\cal M$ such that $a+b=1$ in the
spirit of corresponding position-dependent mass problems~\cite{RosasOrtiz2020}. 
Similar forms of the Hamiltonian as in Eq.~\eqref{postulate.eqn} have been used in Refs.~\cite{Fazio} and~\cite{Joy} for the Dirac equation with a \(z\)-dependent Fermi velocity. 
In fact Eq.~\eqref{postulate.eqn} at the expense of introducing a $z$-dependent matrix ${\cal M}$, appears like a Dirac equation for a locally upright Dirac cone. 
In that respect it can be considered as a solid-state physicist way of introducing Einstein's vielbeins (or frame fields)~\cite{Yepez2011}. 
As we will see in Eq.~\eqref{localJz.eqn}, this transformation allows to write current densities that in the new (local) basis looks like corresponding 
expressions for upright Dirac cones. Therefore this important equation secretly has a frame field content. 

Using our representation of the Hamiltonian in Eq.~\eqref{postulate.eqn} the continuity equation is
\begin{equation}\label{Eq05}
\Delta\cdot{\bf J} + {\partial\rho\over\partial t} = 0,
\end{equation}  
where the probability density, \(\rho\), and the probability current density \({\bf J}\) are
\begin{equation}\label{Eq06}
\begin{split}
\rho & =  \Psi^\dagger \Psi, \\ 
{\bf J} & = { v}_F \big[\Psi^\dagger\sigma_x\Psi\big]{\bf e}_x + {v}_F \big[\Psi^\dagger\big(\sigma_z + \zeta(z){\sigma_0}\big)\Psi\big]{\bf e}_z,
\end{split}
\end{equation}  
where \(\Psi\) is the two-component (spinor) wave function, and \({\bf e}_x\) and \({\bf e}_z\) are unit vectors in the \(x\)- and \(z\)-directions, respectively.  
In the general case, \(\Psi\), \(\rho\), and \({\bf J}\) are functions of both position and time. However, for the sake of brevity, this dependence is not explicitly shown in the equations above. The Hamiltonian in Eq.~\eqref{postulate.eqn}, the continuity equation in Eq.~\eqref{Eq05}, and the expressions for \(\rho\) and \({\bf J}\) in Eq.~\eqref{Eq06} are all mutually consistent. Furthermore the Hamiltonian~\eqref{postulate.eqn} is Hermitian in the sense that 
$   \langle\Psi|H|\Phi\rangle=\langle\Phi|H|\Psi\rangle $~\cite{Witten2016}. 

The specific case where \(\zeta\) is constant has been thoroughly studied in our previous work~\cite{AlMarzoog}. That study examined the behavior of massless Dirac particles as they cross the boundary between two materials with different tilt parameters, the occurrence of Klein tunneling, and the spin rotation angle of these particles after crossing the interface. In the present work, we extend this investigation to the case where \(\zeta\) varies as a function of \(z\).  

\begin{figure}[t!]  
\begin{center}  
\includegraphics[width=0.23\textwidth]{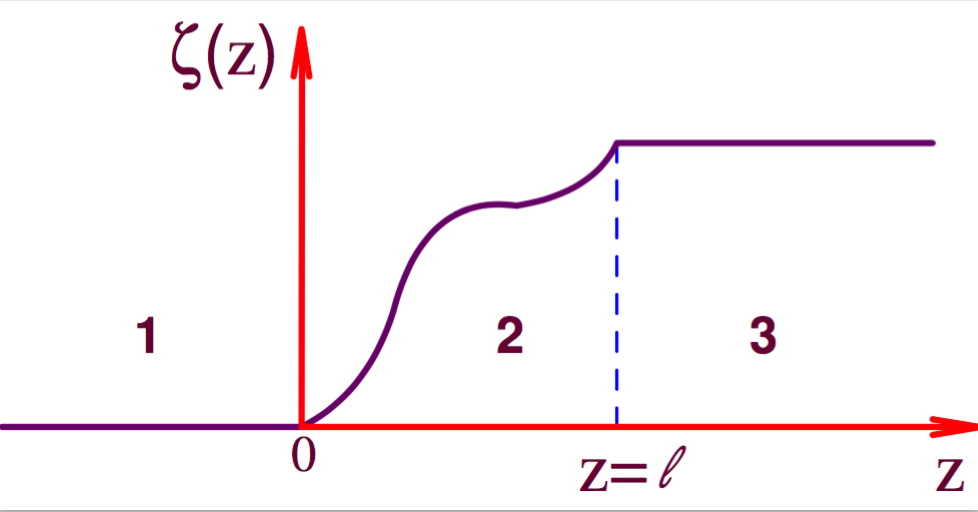}  
\includegraphics[width=0.23\textwidth]{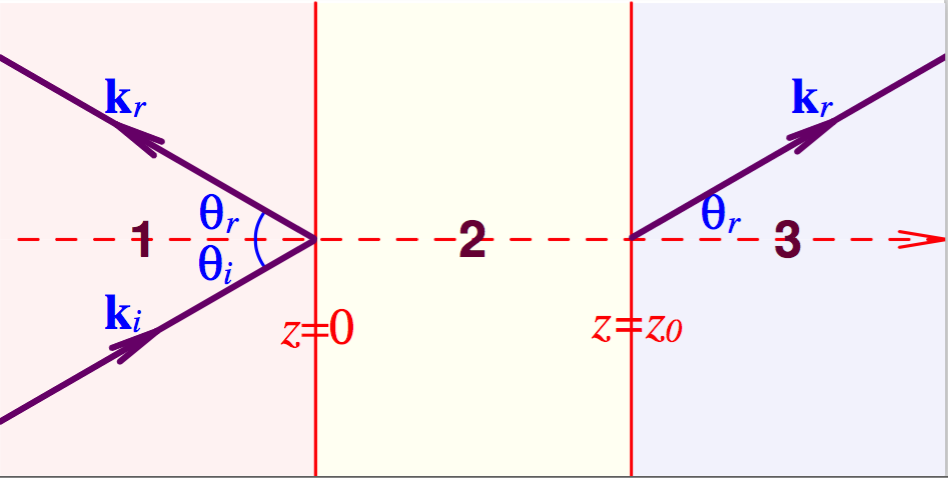}  
\end{center}  
\caption{(Left) Schematic illustration of a heterojunction composed of three materials with different tilt parameters. In the outer regions, the tilting parameter, \(\zeta\) 
is constant, while in the middle layer it varies as an arbitrary function of \(z\). The tilting parameter of the middle layer, \(\zeta(z)\), smoothly varies and
takes the constant values \(\zeta_1=0\) and \(\zeta_3\ne 0\) at the boundaries \(z=0\) and \(z=\ell\), respectively. (Right) Incident, reflected and transmitted electron waves. 
Note that the indicated angles belong to wave vectors that in the presence of tilt does not necessarily coincide with the direction of group velocity.}
\label{FigSchematic}
\end{figure}  

\section{Transmission Probability and Klein Tunneling\label{Sec03}}  

Using the Hamiltonian and the associated continuity equation for tilted Dirac cones with spatially varying tilt parameters, we can explore the 
transport in heterostructure composed schematically depicted in Fig.~\ref{FigSchematic}. In this figure, three regions labeled $1,3$ are separated by a region labeled $2$ within which
the tilt varies in space. The corresponding tilts in general can be \(\zeta_1\), \(\zeta_2(z)\), and \(\zeta_3\). 
The middle layer has a thickness of \(\ell\), while the outer layers extend infinitely.  
In the setup we have studied, region \(1\) is assumed to have no tilt, i.e. $\zeta_1=0$ where the Dirac fermions represented by upright Dirac cones. 
In region \(2\), the tilt continuously varies starting from $\zeta_1=0$ in its left end and continuously reached $\zeta_3$ in its right end. The shape of
variations can be any continuous function $\zeta_2(z)$ schematically displaced in Fig.~\ref{FigSchematic}.

The Schr\"odinger eigenvalue equation, which governs the behavior of massless Dirac fermions in the various regions of the setup depicted in Fig.~\ref{FigSchematic}, is expressed in its general form as  
\begin{equation}\label{Eq07}  
{\cal H} u(x,z) = E u(x,z),  
\end{equation}  
where \(u(x,z)\) represents the eigenfunction of \({\cal H}\) corresponding to the energy eigenvalue \(E\). 
The \(z\)-axis is considered to be normal to the interface of the media. In this configuration, due to the translation invariance in the $x$ direction, 
the \(k_x\) is conserved, and \(u(x,z)\) will be of the following form
\begin{equation}\label{Eq08}  
u(x,z)=\psi(z)e^{ik_x x}.  
\end{equation}  
For this class of wave functions, the Schr\"odinger eigenvalue equation for the middle region simplifies to  
\begin{equation}\label{Eq09}  
{\cal M}(z) \sigma_z {d \over d z} {\cal M}(z)\psi(z) =  i \bigg(\frac{E}{v_F} - k_x \sigma_x \bigg)\psi(z).  
\end{equation}  
Defining the new spinor $\varphi(z)$ by
\begin{equation}
\varphi(z) = {\cal M}(z) \psi(z),  
\label{newspinor.eqn}  
\end{equation}  
Eq.~\eqref{Eq09} becomes
\begin{equation}\label{Eq11}\begin{split}  
{\cal M}^2(z) \sigma_z {d\over dz} \varphi(z) = i \bigg[\frac{E}{v_F} - k_x {\cal M}(z)\sigma_x {\cal M}^{-1}(z)\bigg]\varphi(z).  
\end{split}  
\end{equation}  

To develop some intuition to the structure of solutions, first consider the normal incidence of massless Dirac fermions from region \(1\) onto the interface. 
In this case, the particles travel along the \(z\)-axis, so \(k_x = 0\). Furthermore a complete scattering state can be considered by 
setting the spinor part of the wave function \(\varphi(z)\) and consequently \(\psi(z)\) to have only one non-zero component, i.e. their
spinor part is proportional to either $[1~~0]^T$ or $[0~~1]^T$. For example consider the first case where $\varphi^T(z)=(1~~0)\varphi_+(z)$ which 
corresponds to the boundary conditions  
\begin{equation}\label{Eq12}  
\lim_{z\to - \infty} \zeta(z) = 0, \quad {\rm and} \quad \psi_{in} = \psi(z\to -\infty) = e^{i k_z z},  
\end{equation}  
Eq.~\eqref{Eq11} reduces to  
\begin{equation}\label{Eq13}  
{\cal M}_{11}^2(z) {d\over dz} \varphi_+(z) = i \frac{E}{v_F} \varphi_+(z),  
\end{equation}  
where \({\cal M}_{ij}(z)\) represents the element in row \(i\) and column \(j\) of the matrix \({\cal M}\). The solution to this equation, under the boundary conditions of Eq.~\eqref{Eq12}, is  
\begin{equation}\label{Eq14}  
\varphi_+(z) = e^{i \alpha_+(z)},  
\end{equation}  
where  
\[\alpha_+(z)= -\frac{E}{v_F} \int_{-\infty}^z \frac{dz}{{\cal M}_{11}^2(z)}.\]  
Finally, using Eq.~\eqref{newspinor.eqn}, \(\psi_+(z)\) is determined as  
\begin{equation}\label{Eq15}  
\psi_+(z) = \frac{1}{{\cal M}_{11}(z)} e^{i \alpha_+(z)}
\begin{bmatrix}1\\0 \end{bmatrix}.  
\end{equation}  

Eq.~\eqref{Eq15} shows that the \(z\)-component of the probability current density, \(J_z\), remains constant, signifying the occurrence of Klein tunneling. In this scenario, which corresponds to normal incidence at the interface, there is a perfect transmission probability, allowing all particles to traverse the central slab and enter the third medium.  

If instead the boundary conditions are  
\begin{equation}\label{Eq16}  
\lim_{z\to + \infty} \zeta(z) = 0, \quad {\rm and} \quad \psi_{in}=\psi(z\to +\infty) = e^{- i k_z z},  
\end{equation}  
then in the above equations, \({\cal M}_{11}(z)\) should be replaced by \({\cal M}_{22}(z)\) and the
corresponding $\sigma_z$ eigen-spinor must be replaces by $[0~~1]^T$. 

Next, let us consider a more general case. As illustrated in the right panel of Fig.~\ref{FigSchematic}, we assume that massless Dirac fermions travel through medium \(1\) 
and strike the interface with the middle slab at an arbitrary angle. 
Our goal is to calculate the reflection and transmission probabilities for these states 
as they traverse the intermediate region.  

Referring to the right panel of Fig.~\ref{FigSchematic}, the wave function of the system in the outer regions is  
\begin{equation}\label{Eq17}  
\psi(z) = \left\{\begin{split}  
{1 \over t} & \psi_i(z) + {r \over t} \psi_r (z),\qquad\qquad  z<0,\\  
& \psi_t(z),\qquad \qquad\qquad\qquad z>z_0,  
\end{split}\right.  
\end{equation}  
where  
\begin{equation}\label{Eq18}  
\begin{split}  
\psi_i(z) &= \begin{bmatrix} \cos{\theta_i \over 2} \\ \sin{\theta_i \over 2} \end{bmatrix} e^{i k_{iz} z}, \\  
\psi_r(z) &= \begin{bmatrix} -\sin{\theta_i \over 2} \\ \cos{\theta_i \over 2} \end{bmatrix} e^{-i k_{rz} z}, \\  
\psi_t(z) &= \begin{bmatrix} \cos{\theta_t \over 2} \\ \sin{\theta_t \over 2} \end{bmatrix} e^{i k_{tz} z},  
\end{split}  
\end{equation}  
where subscripts $i$, $r$ and $t$, refer to incident, reflected and transmitted waves, respectively. 
The coefficients \(r\) and \(t\) represent the reflection and transmission amplitudes. The angles \(\theta_i\), \(\theta_r\), and \(\theta_t\), termed the angles of incidence, reflection, and transmission, respectively~\footnote{Note that for the upright Dirac cones, the direction of wave vectors
are the same as group velocities, whereas in the tilted Dirac cone case they are not the same. We choose to parameterize the angular variables
refering to the wave vectors~\cite{AlMarzoog}.} as shown in Fig.~\ref{Fig02}. 
Correspondingly, the wave vectors associated with the incident, reflected, and transmitted waves are denoted as \({\bf k}_i\), \({\bf k}_r\), and \({\bf k}_t\). 

The components of these wave vectors in the \(z\)-direction (normal to the interface) are \(k_{iz}\), \(k_{rz}\), and \(k_{tz}\), while their components in the \(x\)-direction 
are \(k_{ix}\), \(k_{rx}\), and \(k_{tx}\). Due to translational invariance along $x$ direction, all the $x$-components of the wave vectors are equal and therefore
they are all specified with a single vlaue $k_x$
\begin{equation}  
k_x = k_{ix} = k_{rx} = k_{tx}.  
\end{equation} 
It is crucial to note that the value of $z$-components of wave vectors adjust themselves to satisfy the boundary conditions 
that ensure continuity equation. The value of \(k_{tz}\) depends on the tilting parameter in region~3. 
Moreover, the energy $E$ is also converted in all regions and in region~1 where there is no tilting, we have:  
\begin{equation}\label{Eq19}  
E = \sqrt{k_{iz}^2 + k_x^2} = \sqrt{k_{rz}^2 + k_x^2},  
\end{equation}  
allowing us to easily determine \(k_{iz}\) and \(k_{rz}\).  

To compute \(k_{tz}\), we note that in region~3, the Schr\"odinger eigenvalue equation from the Hamiltonian in Eq.~\eqref{Eq01} is
\begin{equation}\label{Eq20}  
\big( k_{tz} \sigma_z + k_x \sigma_x \big) \psi_t(z) = \big(E - \zeta_3 k_{tz} \big) \psi_t(z).  
\end{equation}  
The eigenvalues of the operator on the left-hand side are \(\pm \sqrt{k_x^2 + k_{tz}^2}\). Since we are considering waves propagating to the right (\(k_{tz} > 0\)), we choose the positive eigenvalue. Setting this eigenvalue equal to \(E - \zeta_3 k_{tz}\), we obtain  
\begin{equation}\label{Eq21}  
E = \zeta_3 k_{tz} + \sqrt{k_x^2 + k_{tz}^2},  
\end{equation}  
which is an implicit equation for $k_{tz}$ that can be solved for \(k_{tz}\).  
With the \( z \)-components of the incident, reflected, and transmitted wave vectors, along with the known angles of incidence, reflection, and transmission, 
the wave functions \(\psi_i(z)\), \(\psi_r(z)\), and \(\psi_t(z)\) are uniquely determined.  

To fully characterize the wave function of the system outside the central slab, it is crucial to determine the amplitudes \(r\) and \(t\) defined 
in Eq.~\eqref{Eq17}. This requires a detailed analysis of the wave function behavior within the central region. By examining this wave function and applying the 
continuity conditions, the values of \(r\) and \(t\) can be precisely determined.  
To achieve this, we utilize Schrd\"oinger's equation for a fixed energy $E$, as given by Eq.~\eqref{Eq11} for the central slab region. 
In this region, the wave function \(\varphi(z)\) is represented as a two-component spinor
\begin{equation}\label{Eq22}  
\varphi(z) = \begin{bmatrix}  
\varphi_1(z) \\  
\varphi_2(z)  
\end{bmatrix}.  
\end{equation}  
In the middle slab (\(0 \leq z \leq \ell\)), Eq.~\eqref{Eq11} can be separated into two coupled first-order differential equations governing \(\varphi_1(z)\) and \(\varphi_2(z)\):  
\begin{equation}\label{Eq23}  
\begin{split}  
A(z) \frac{d}{dz} \varphi_1(z) & - i B(z) \varphi_1(z) + i k_x \varphi_2(z) = 0, \\  
A(z) \frac{d}{dz} \varphi_2(z) & + i C(z) \varphi_2(z) - i k_x \varphi_1(z) = 0,  
\end{split}  
\end{equation}  
where the coefficients \(A(z)\), \(B(z)\), and \(C(z)\) are defined as:  
\begin{equation}\label{Eq24}  
\begin{split}  
A(z) &= {\cal M}_{11}(z) {\cal M}_{22}(z), \\  
B(z) &= \left(\frac{E}{v_F}\right) \frac{{\cal M}_{22}(z)}{{\cal M}_{11}(z)}, \\  
C(z) &= \left(\frac{E}{v_F}\right) \frac{{\cal M}_{11}(z)}{{\cal M}_{22}(z)}.  
\end{split}  
\end{equation}  

For arbitrary variable functions of \(\zeta(z)\) satisfying the appropriate boundary conditions (as illustrated in Fig.~\ref{FigSchematic}), 
these coupled first-order differential equations can be numerically solved to desired accuracy. The boundary conditions are derived from the continuity of the \(z\)-component of the probability current density at the interfaces. Using the expression for \({\bf J}\) from Eq.~\eqref{Eq06}, we can express the \(z\)-component of the current density as:  
\begin{equation}
\begin{split}  
J_z &= {v}_F \psi^\dagger(z) \big(\sigma_z + \zeta(z) {\sigma_0}\big) \psi(z) \\  
&= {v}_F \psi^\dagger(z) {\cal M}(z) \sigma_z {\cal M}(z) \psi(z) \\  
&= {v}_F \varphi^\dagger(z) \sigma_z \varphi(z).  
\label{localJz.eqn}
\end{split}  
\end{equation}  
As pointed out below Eq.~\eqref{postulate.eqn}, the last line looks like the current density for a locally upright Dirac cone. 
Therefore the new spinor $\varphi(z)$ defined by Eq.~\eqref{newspinor.eqn} can be interpreted as a spinor in a locally Minkowski spacetime. 
As such, the matrix ${\cal M}(z)$ defined in Eq.~\eqref{postulate.eqn} encapsulates a transformation to a locally flat Minkowski spacetime. 

To proceed with the question of boundary conditions, since \(J_z = { v}_F \varphi^\dagger(z) \sigma_z \varphi(z)\), it follows that \(\sqrt{{v}_F} \varphi(z)\) 
is continuous across all regions, including at the interfaces. In particular, at the interface between regions~2 and~3 (\(z = \ell\)), this continuity implies:  
\begin{equation}\label{Eq26}  
\sqrt{{v}_{2F}}  
\begin{bmatrix}  
\varphi_1(\ell) \\  
\varphi_2(\ell)  
\end{bmatrix} = \sqrt{{v}_{3F}}  
\begin{bmatrix}  
\cos(\theta_t / 2) \\  
\sin(\theta_t / 2)  
\end{bmatrix},  
\end{equation}  
where \({v}_{2F}\) and \({v}_{3F}\) are the Fermi velocities in regions~2 and~3, respectively.  
This boundary condition is essential to obtain unique numerical solutions \(\varphi_1(z)\) and \(\varphi_2(z)\) of the coupled differential equaitons~\eqref{Eq23}. 
In this work we will assume that all regions have the same Fermi velocity $v_F$. 

By applying the continuity condition at the boundary \(z=0\), the following expression is derived:  
\begin{equation}\label{Eq27}  
\frac{r}{t} \begin{bmatrix}  
\sin\left(\frac{\theta_i}{2}\right) \\  
\cos\left(\frac{\theta_i}{2}\right)  
\end{bmatrix} + \frac{1}{t} \begin{bmatrix}  
\cos\left(\frac{\theta_i}{2}\right) \\  
\sin\left(\frac{\theta_i}{2}\right)  
\end{bmatrix} =  
\begin{bmatrix}  
\varphi_1(0) \\  
\varphi_2(0)  
\end{bmatrix}.  
\end{equation}  
This equation consists of two coupled algebraic equations the solution of which readily gives the amplitudes of reflection (\(r\)) and transmission (\(t\)),
\begin{equation}\label{Eq28}  
\begin{split}  
\frac{1}{t} &= \frac{\varphi_1(0)\cos\left(\frac{\theta_i}{2}\right) - \varphi_2(0)\sin\left(\frac{\theta_i}{2}\right)}{\cos(\theta_i)}, \\  
\frac{r}{t} &= \frac{\varphi_1(0)\sin\left(\frac{\theta_i}{2}\right) - \varphi_2(0)\cos\left(\frac{\theta_i}{2}\right)}{\cos(\theta_i)}.  
\end{split}  
\end{equation}  
With these amplitudes, the reflection and transmission probabilities $R$ and $T$ can be readily computed:
\begin{equation}\label{Eq29}  
\begin{split}  
R &= \frac{J_r}{J_{\text{in}}} = |r|^2, \\  
T &= \frac{J_t}{J_{\text{in}}} = |t|^2 \left(\frac{\varphi_1^2(0) - \varphi_2^2(0)}{\cos^2(\theta_i)}\right), 
\end{split}  
\end{equation}  
Needless to say, they satisfy $R+T=1$. 


\begin{figure}[t!]  
\begin{center}  
\includegraphics[width=0.40\textwidth]{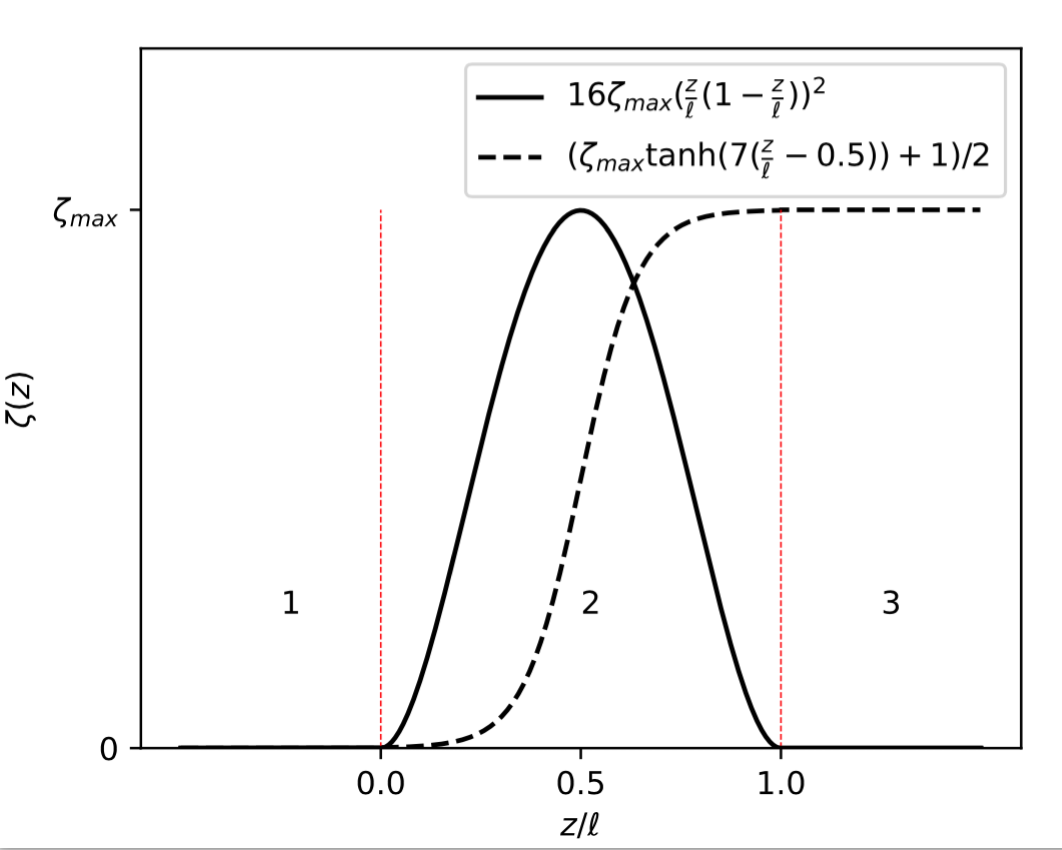}  
\end{center}  
\caption{  
Two distinct tilting functions, \(\zeta_b(z)\) (solid lines) and \(\zeta_s(z)\) (dashed lines) describing a bump and sigmoid shaped profile of tilting in 
the middle layer of the heterojunction used in this study. The functional forms are indicated in figure and Eqs.~\eqref{Eqzetab} and~\eqref{Eqzetas} represent
situations where stronger tilt is concentrated in the middle and right side of the junction, respectively. 
}  
\label{Fig03}  
\end{figure}  

\section{Results and Discussion~\label{Sec04}}

\begin{figure*} 
\begin{center}
\includegraphics[width=1.00\textwidth]{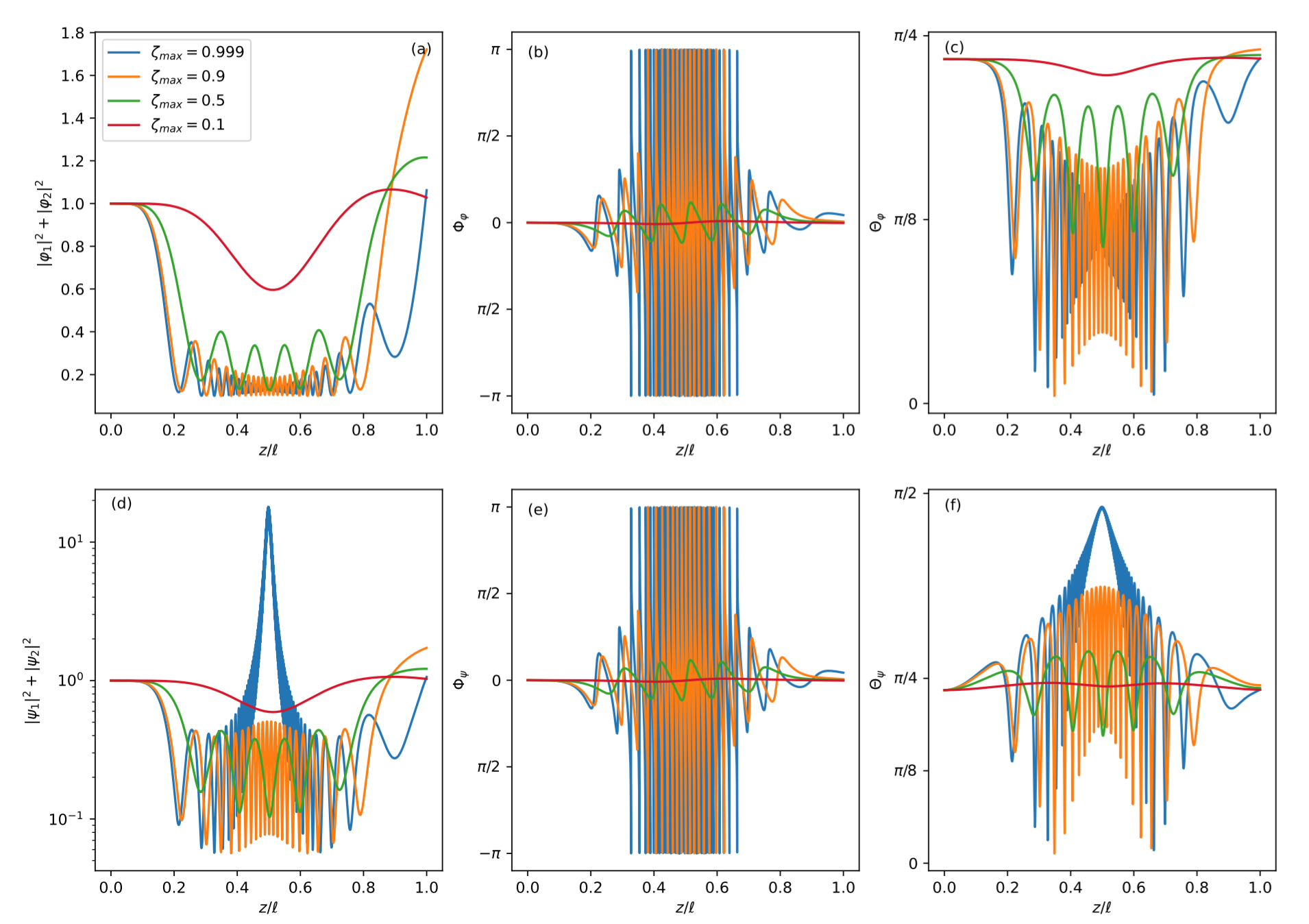}  
\end{center}
\caption{  
Numerical solutions of spinor amplitude, polar angle and azimuthal angle in local frame (first row) and substrate frame (second row)
as functions of \( z / \ell \) for \(\theta_i = \frac{\pi}{2} - 0.1\), \(\ell = 50\), $E=1\times \hbar v_F/\ell$, 
and various tilt parameters \(\zeta_{\text{max}} = 0.1, 0.5, 0.9, 0.999\). 
 \( |\varphi(z)| \), \( \Theta_\varphi(z) \), and \( \Phi_\varphi(z) \) are given in panels (a), (b), and (c). 
\( |\psi(z)| \), \( \Theta_\psi(z) \), and \( \Phi_\psi(z) \) are given in panels (d), (e), and (f)
Increasing \(\zeta_{\text{max}}\) leads to an enhanced oscillation frequency across all panels, reminiscent of gravitational redshift. The energy $E=1$ sets
the natural unit of energy and also the unit of length $\lambda_E=\hbar v_F/E$. 
}
\label{Fig-TwoFrames}
\end{figure*}

In this section, we perform a detailed analysis of the solution to the set of differential equations in Eqs.~\eqref{Eq23}. We begin by examining a heterojunction composed 
of three regions where regions $1$ and $3$ have two uniform tilts as shown in Fig.~\ref{Fig03}. The middle layer $2$ has a position-dependent tilt. In our work we have considered
two distinct tilting textures $\zeta_{b}$ and $\zeta_s$ for "bump" and "sigmoid" shaped functions given by 
\begin{equation}
\zeta_b(z) = 16\zeta_{\text{max}} \left( \frac{z}{\ell} \left( 1 - \frac{z}{\ell} \right) \right)^2,
\label{Eqzetab}
\end{equation}
where \( \zeta_{\text{max}} \) is a parameter that we take at most to be $1^-$, 
and \( \ell \) represents the length of the middle layer, measured in units of \(\lambda_E= \hbar v_{F}/E \). 
Note that due to linear energy-momentum scaling, fixing the energy $E$, fixes this natural unit $\lambda_E$ for the length. 
Therefore the length $\ell$ and energy $E$ are not independent variables. Studying the problem at higher (lower) $E$ or smaller (larger) $\lambda_E$ 
means that a fixed value, e.g. $\ell=50$ in units of $\lambda_E$ will correspond to longer (shorter) physical lengths for the variable-tilt region. 
For the sigmoid shaped tilting profile we choose, 
\begin{equation}
\zeta_s(z) = \frac{\zeta_{\text{max}} \tanh\left( 7 \left( \frac{z}{\ell} - 0.5 \right) \right) + 1}{2},
\label{Eqzetas}
\end{equation}
which introduces an alternative tilting profile.
Both functions are displayed in Fig.~\ref{Fig03} where \( \zeta_b(z) \) is shown by solid lines and \( \zeta_s(z) \) by dashed line. 

Next, we solve the set of Eqs.~\eqref{Eq23} by considering an incoming electron with a given wave vector \( \mathbf{k_i} = (k_{ix}, k_{iz}) = (\cos(\theta_i), \sin(\theta_i)) \). It is clear that \( \theta_t = \theta_i \), and therefore, \( \Big[ \begin{array}{c} \varphi_1(\ell) \\ \varphi_2(\ell) \end{array}\Big] \) will be determined according to Eq.~\eqref{Eq26}. This provides the boundary conditions for the equations at \( z = \ell \) and allows us to numerically find the solutions of these equations.

Due to the complex structure of the spinors \( \varphi_1(z) \) and \( \varphi_2(z) \), the final solutions of the system are parameterized as
\begin{equation}
\Big[ \begin{array}{c} \varphi_1(z) \\ \varphi_2(z) \end{array}\Big] = |\varphi(z)|  \left[ \begin{array}{c} \cos(\Theta(z)/2) \\ \sin(\Theta(z)/2) e^{i\Phi(z)} \end{array} \right],
\end{equation}
where \( |\varphi(z)|^2 = |\varphi_1(z)|^2 + |\varphi_2(z)|^2 \), where $\Theta(z) \in [0, \pi)$
and \( \Phi(z) \in [0, 2\pi) \) are usual spherical angles which for a spinor are given by
\begin{eqnarray*}
&&\Theta(z) = \tan^{-1}\left( \frac{|\varphi_2(z)|}{|\varphi_1(z)|} \right),\\
&&\Phi(z) = \tan^{-1}\left(\frac{\Im(\varphi_2)}{\Re(\varphi_2)}\right) - \tan^{-1}\left(\frac{\Im(\varphi_1)}{\Re(\varphi_1)}\right).
\end{eqnarray*}
The angles $\Theta$ and $\Phi$ can be regarded as the polar and azimuthal angles of the spinors in locally flat Minkowski spacetime. 
In the following, we present numerical results for \( |\varphi(z)| \), \( \Theta(z) \), and \( \Phi(z) \) obtained using the 4th-order Runge-Kutta method. 

We choose the parameter of the incident wave \( \theta_i = \frac{\pi}{2} - 0.1 \) and take the length of the middle layer to be \( \ell = 50 \) (in units of $\lambda_E$).  
The corresponding numerical solutions are presented in Fig.~\ref{Fig-TwoFrames}. The first row of panels, labeled (a), (b), and (c), displays 
\( |\varphi(z)| \), \( \Theta_\varphi(z) \), and \( \Phi_\varphi(z) \) as functions of \( z / \ell \) for four different values of the tilt heights: \( \zeta_{\text{max}} = 0.1, 0.5, 0.9, 0.999 \) which specifies the height of the bump. As can be seen upon increasing the magnitude of the bump, more oscillations are induced to all of these
quantities as a function of the dimensionless length \( z / \ell \). Another salient feature of all the figures of the top row that represent the spinor components 
and the spherical angles as a function of position is that for large enough $\zeta_{\rm max}$ where enough oscillations are formed, the wave length of the oscillations
are smaller in the center of the bump whereas the wave lengths are larger (i.e. red-shifted) away from the bump. 
Note that the normalization of $\varphi$ is such that the continuity equation holds. The physically observable transmission coefficients are calculated based on the 
ratios of the amplitudes at beginning and end of the middle region, and this normalization does not pose any issues. 

As pointed out the spinor $\varphi$ represents the wave function in a local frame. It is useful to transform it back to the original frame (that can be dubbed
substrate frame to emphasize its solid state context) by the defining relation \( \psi = {\cal M}^{-1} \varphi \). 
The results are shown in second row as panels (d), (e), and (f) of Fig.~\ref{Fig-TwoFrames}. For clarity, we use the notations 
\( \Theta_\varphi \) and \( \Phi_\varphi \) for the spherical angles of \( \varphi \), and \( \Theta_\psi \) and \( \Phi_\psi \) for those of \( \psi \). 

Notably, in both frames the oscillation are faster in the regions where the local value of $\zeta$ is larger which is reminiscent of "gravitational" red-shift pattern. 
As can be seen this behavior manifests itself in both local frame ($\varphi$) and substrate frame ($\psi$). 
In particular, the oscillation patterns of the azimuthal angle $\Phi_\psi$ and $\Phi_\varphi$ are quite similar. Note that the azimuthal angles
are with respect to $z$-axis that also denotes the direction of transport. Therefore the azimuthal angles both represent transverse precession in the local frame and 
and the substrate frames.

The behavior of polar angle that represents the orientation of the spinor with respect to the motion direction is however quite different
which can be seen from the comparison of panels (c) and (f). 
Let us begin by noting the similarities in the wavelength of the oscillations: both panels clearly share the red-shift behavior. 
In the substrate frame, as can be seen in panel (f), by increasing the bump strength $\zeta_{\rm max}$ in the center of the middle region where $\zeta$ 
can reach its $\zeta_{\rm max}$ value, the average value around which the oscillations is taking place is lifted upward. For $\zeta_{\rm max}=0.999$ (blue curve) that is very 
close to the "horizon" value of $\zeta=1$ this enhancement of the $\Theta$ and its tendency to "equator" value of $\pi/2$ is more manifest in panel (f).  
This means that in the substrate frame first of all the horizon is indicated as locking of the polar angle to equatorial value, and secondly the nodding amplitude
is suppressed more and more. The suppression of the nodding amplitude is also seen in panel (c) that represents the polar angle in local frame. 
However, in this frame the saturation value of the polar angle is $\approx\pi/8$ which is much smaller than the equatorial value. 

A comparison can also be made between panels (a) and (d) that represent the total probability constructed from both components of the spinors in local and 
substrate frames, respectively. Again as can be seen both frames display clearly the red-shift pattern in their wave lengths. In the local frame as can be
seen in panel (a), the average value around which the oscillations are taking place approaches a fixed value at the center where $\zeta$ has its maximum value. 
The approach to horizon is remarkably reflected in the behavior of the probability $|\psi|^2$ in panel (d) as a strong peak (note the logarithmic scale in panel d). 
Comparing panel (d) against (a), can be seen that the approach to the critical value $\zeta=1$ can be diagnosed in the substrate coordinate in which the solid-state 
measurements are done. Local frame that moves along with the electron wave is only a mathematically convenient frame for the computation. 

Combining the information from $\Theta$ and the probability if one has a local probe of pseudospin, the enhancement of the local $\zeta$ can be inferred from both 
the noting amplitude as well as the strength of the signal that will be directly controlled by the square of wave function. 

\subsection{Analytical result for the position-dependent oscillations}
Can one understand the above numerical results analytically? It turns out that the answer is yes, and a very illuminating physical picture emerges. 
To address the physical origin of position-dependent or equivalently $\zeta(z)$-dependent oscillations, we focus on a small enough interval \( \Delta z \) of \( z \) 
where the function \( \zeta(z) \) can be approximated as constant. Within this interval, the Schr\"odinger equation \eqref{Eq09} for \( \varphi \), 
derived from Eq.~\eqref{Eq22}, takes the form:
\begin{equation}\label{E31}
i{\cal M}^2 \sigma_z \frac{d}{dz} \varphi(z) =  
\big( -E \sigma_0 + k_x {\cal M} \sigma_x {\cal M}^{-1} \big) \varphi(z),
\end{equation}  
where \( \sigma_0 \) is the \( 2 \times 2 \) identity matrix. 
Now in the spirit of WKB approximation, we seek solutions of the form \( \varphi(z) \propto e^{-ik_z z} \) where $k_z(z)$ depends on $z$ and hence
local value of $\zeta$. Substituting this ansatz into Eq.~\eqref{E31}, we obtain two eigenvalues for \( k_z \) denoted by \( k_z^+ \) and \( k_z^- \) which are given by:
\begin{equation}\label{Eq33}
k_z^\pm = \frac{-\zeta E \pm \sqrt{E^2 - (1 - \zeta^2)k_x^2}}{1 - \zeta^2}.
\end{equation}  
By transforming the Pauli matrix to local frame, namely \( \sigma_{x,z}' = {\cal M} \sigma_{x,z} {\cal M}^{-1} \) we have $\sigma_z'=\sigma_z$ and
\( \{\sigma_x', \sigma_z'\} = 0 \) and \( (\sigma_x')^2 = 1 \). As a result, the eigenvalues \( k_z^+ \) and \( k_z^- \) in Eq.~\eqref{E31} are identical to those of:
\begin{equation}\label{E32}
i{\cal M}^2 \sigma_z \frac{d}{dz} \psi(z) =  
\big( -E \sigma_0 + k_x \sigma_x \big) \psi(z),
\end{equation}  
where the spinors in the two frames are related by \( \varphi = {\cal M} \psi \).  

Let the eigenfunctions of Eq.~\eqref{E32} be \( \psi^+ \) and \( \psi^- \), and those of Eq.~\eqref{E31} be \( \varphi^+ \) and \( \varphi^- \). The relationship between these eigenfunctions is given by
\begin{equation}\label{E33}
\langle \varphi^+ | \varphi^- \rangle = 2 \zeta (\psi_1^+)^* \psi_2^-,
\end{equation}  
where the subscripts $1,2$ in the right hand side indicate the first and second spinor {\em components}. 
This relation shows that the eigenfunctions \( \varphi^+ \) and \( \varphi^- \) in the local frame are not orthogonal and the
amount of non-orthogonality is determined controlled by the amount of tilt parameter $\zeta$ vanishing for $\zeta\to 0$. 

If the initial state of the system is chosen as a linear combination of the eigenbasis, i.e., \( \varphi(z=0) = \alpha_+ \varphi^+ + \alpha_- \varphi^- \), the norm of \( \varphi(z) \) varies with \( z \) as:
\begin{equation}\label{E34}
|\varphi(z)|^2 = 2 \zeta \alpha_+ \alpha_- \sin(\Delta k_z z),
\end{equation}  
where \( \Delta k_z = k_z^+ - k_z^- \). This result demonstrates that \( |\varphi(z)|^2 \) oscillates periodically as a function of \( z \), and for a sufficiently large length, it may even return to its initial value.
\begin{figure}[t!]
\begin{center}
\includegraphics[width=0.50\textwidth]{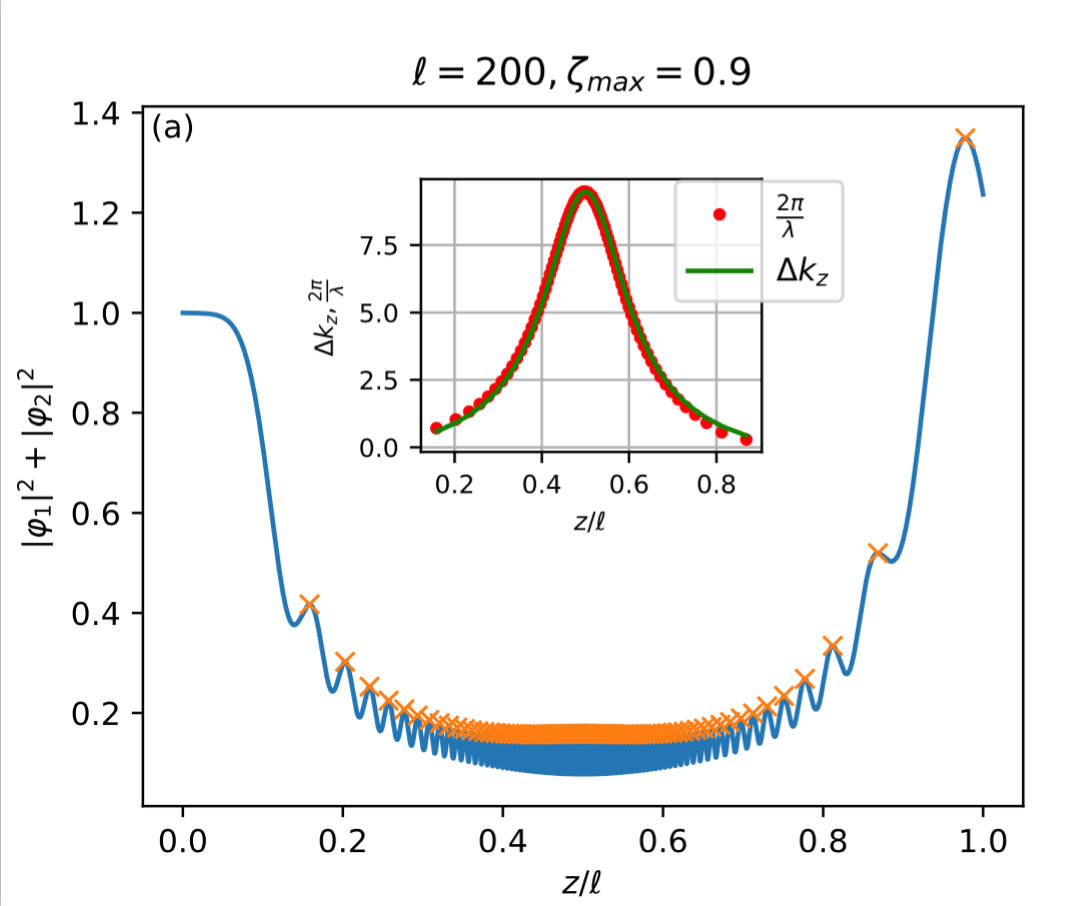}  
\includegraphics[width=0.50\textwidth]{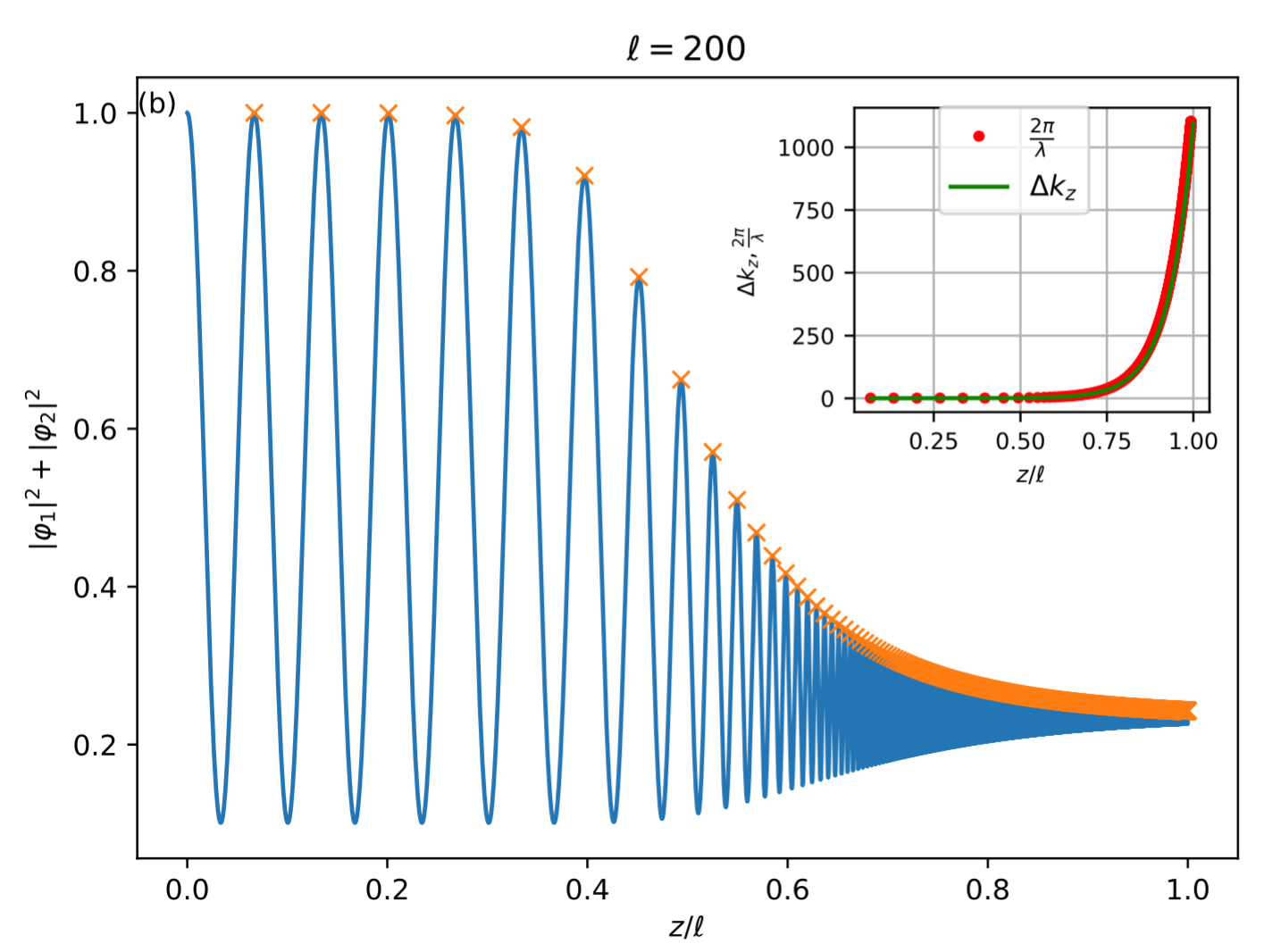}  
\end{center}
\caption{Variation of oscillation wavelengths for the tilting functions \( \zeta_b(z) \) and \( \zeta_s(z) \). The top panel (a) shows the oscillations of \( |\varphi_1|^2 + |\varphi_2|^2 \) for the bump profile \( \zeta_b \) as a function of \( z/\ell \) with \( \ell = 200 \), where the peaks of the oscillations are marked by crosses (\(\times\)). The bottom panel (b) presents the same analysis for the sigmoid profile \( \zeta_s \). Insets in both panels show the calculated wave number \( \Delta k_z(z) \) (red dots), derived from the peak separations, alongside the analytic expression for \( \Delta k_z(z) \) (green curve), demonstrating excellent agreement between the numerical and analytical results.
The value of $k_x$ is such that it correspond to $\theta_i=\pi/2-0.1$. Energy is $E=1$ and $\zeta_{\rm max}=0.9$. 
}
\label{Fig05}
\end{figure}
%
This analysis even allows to extrapolate to neighboring point $z+\Delta z$. The solution in an adjacent interval can be expressed as:  
\begin{equation}\label{E35}
\varphi(z + \Delta z) = e^{i\Lambda(z) \Delta z} \varphi(z),
\end{equation}  
where \( \Lambda(z) = k^+ |\varphi^+\rangle \langle \varphi^+| + k^- |\varphi^-\rangle \langle \varphi^-|\). It is evident that even small changes in 
\( z \) alter the norm of
\( \varphi \).  
Furthermore, it can be shown that the expectation value of \( \sigma_z \) is conserved, i.e., \( \langle \varphi(z) | \sigma_z^' | \varphi(z) \rangle = \text{const.} \), 
since \( \frac{\partial}{\partial z} \langle \varphi(z) | \sigma_z^' | \varphi(z) \rangle = 0 \). This result is consistent with the earlier discussion on 
current density conservation. Algebraically it derives from the fact that sicne both $\sigma_z$ and $\cal M$ are diagonal matrices, the Pauli matrix $\sigma_z$ in both
substrate frame and local frame is the same.
While the norm of \( \varphi \) oscillates with \( z \), it can also acquire a phase factor due to precession around the \( z \)-axis.  

\begin{figure}[t!]
\begin{center}
\includegraphics[width=0.50\textwidth]{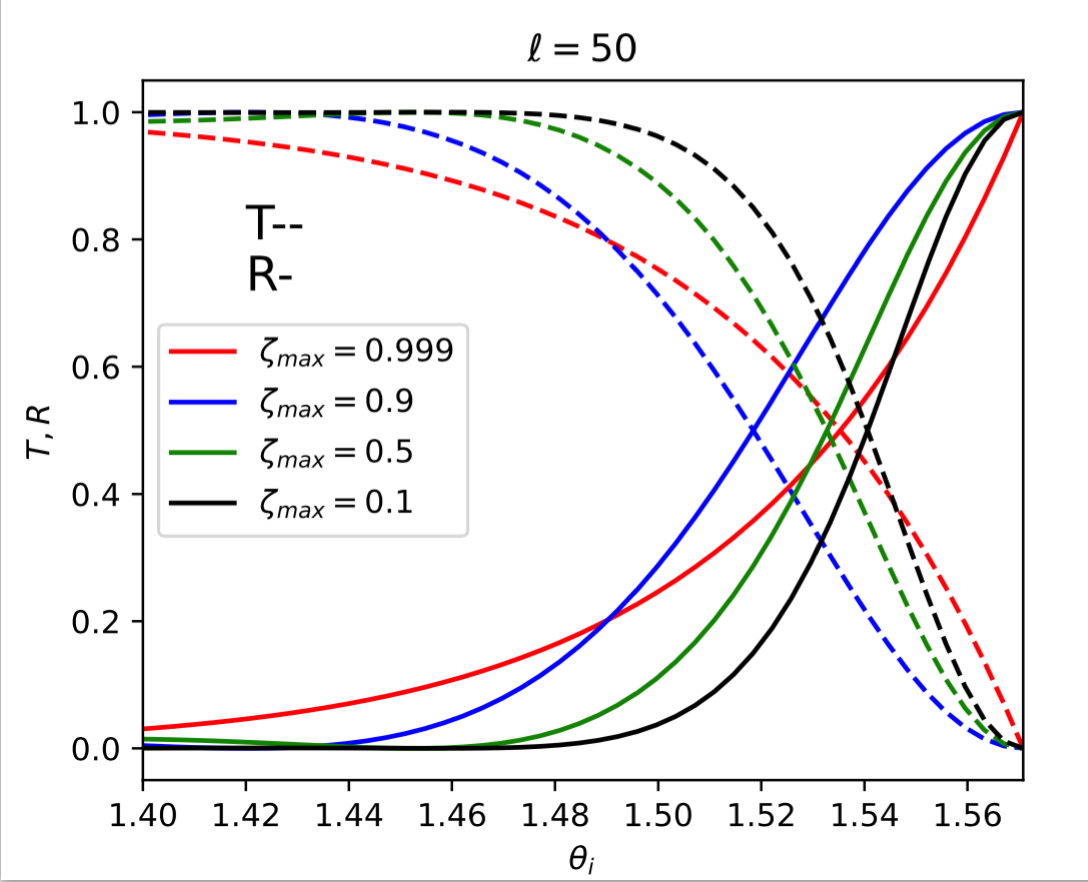}  
\end{center}
\caption{Transmission (\(T\)) and reflection (\(R\)) coefficients as functions of the angle \(\theta_i\) for a fixed middle layer length \(\ell = 50\) for the tilting bump
$\zeta_b$ profile for values of the tilting bump hight parameter \(\zeta_{\text{max}} = 0.1, 0.5, 0.9, 0.999\) as indicated. 
Energy is $E=1\times\hbar v_F/\ell$.
}
\label{Fig06}
\end{figure}

To corroborate the above analytical result with numerical computations, in Fig.~\ref{Fig05} for a middle layer with \( \ell = 200 \) we examine the variation in 
\( |\varphi_1|^2 + |\varphi_2|^2 \) for the two tilting functions \( \zeta_b \) (top panel (a)) and \( \zeta_s \) (bottom panel (b)) 
defined in Eqns.~\eqref{Eqzetab} and~\eqref{Eqzetas} and for the incident wave with $\theta_i=\pi/2-0.1$. 
The oscillation peaks in Fig.~\ref{Fig05} are marked by orange crosses. By locating the positions of adjacent peaks we determined the wavelength 
\( \lambda(z) \) from which the wave number \( \Delta k_z(z) \) is obtained as \( 2\pi / \lambda(z) \). In the insets the wave numbers for both \( \zeta_b \) and \( \zeta_s \) 
are compared with the following analytic expression \( \Delta k_z(z) \):
\begin{equation}
\Delta k_z(z) = \frac{2 \sqrt{1 - (1 - \zeta^2(z))\sin^2(\theta_i)}}{1 - \zeta^2(z)}. 
\label{Deltakz.eqn}
\end{equation} 
As can be seen a very good agreement between this analytic wave numbers and numerical results for both tilting functions
can be achieved.
Panel (b) further indicates that the red-shift behavior is not limited to the bumped tilt texture. 
Such a red-shift behavior can be considered as a probe of the local value of  $\zeta$. 
Our analysis confirms that similar wave numbers can be obtained for other quantities, such as \( \Phi \) and \( \Theta \).  
\begin{figure}[t!]
\begin{center}
\includegraphics[width=0.48\textwidth]{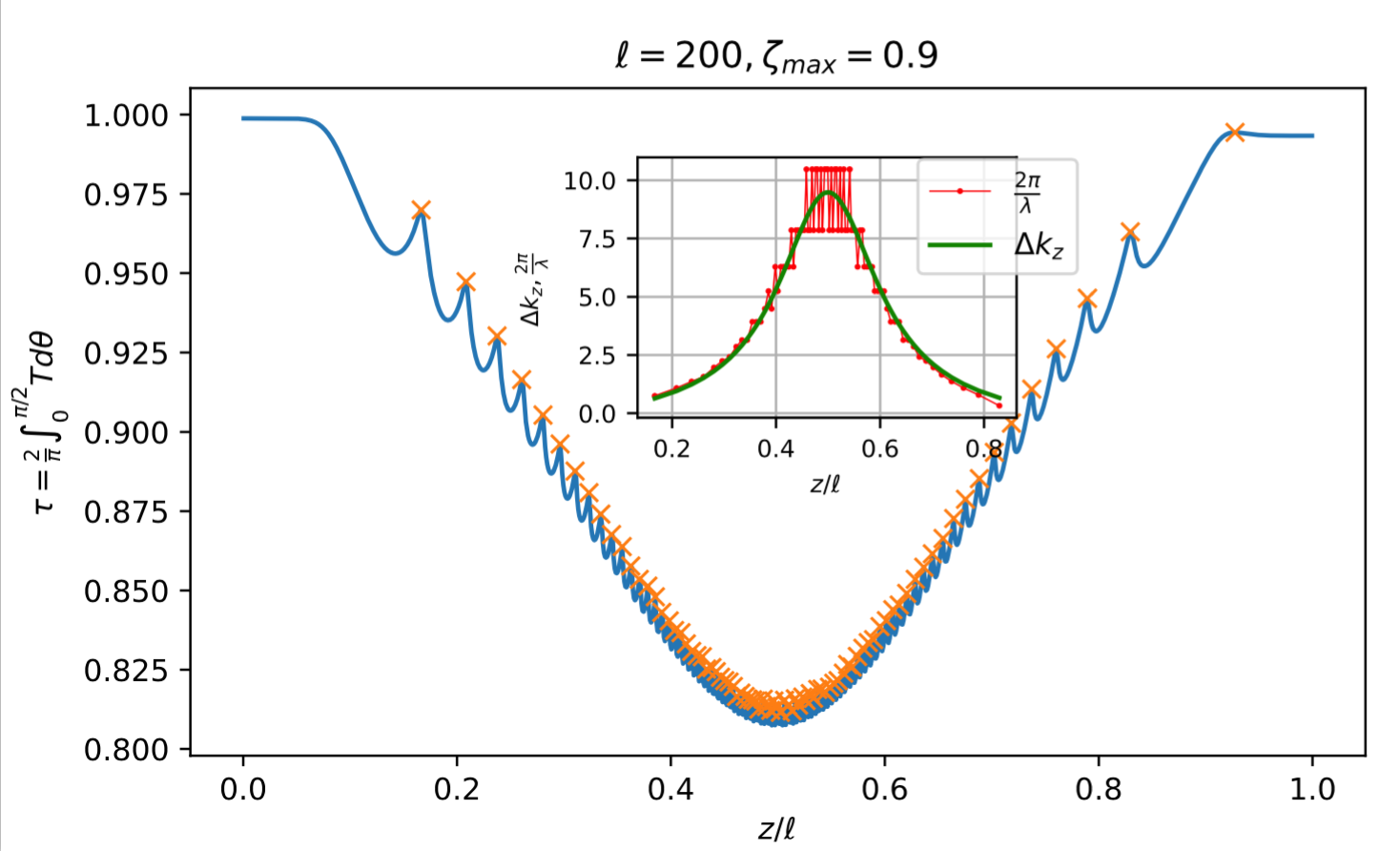}  
\end{center}
\caption{
Integrated transmission \(\tau\) as a function of $z/\ell$ representing the fraction of the tilted region traversed by the incident wave. 
The oscillations in \(\tau\) reflect interference effects caused by wave propagation through the tilted region. The inset shows the wavenumber
\(2\pi / \lambda\) extracted from \(\tau\) which coincides with \(\Delta k_z\), demonstrating a consistent agreement with the wavenumber derived analytically 
(see Fig.~\ref{Fig05}). Energy is $E=1$ and $\ell=200$ in units of $\hbar v_F/E$. } 
\label{FigTransmission}
\end{figure}
\begin{figure}[b!]
\begin{center}
\includegraphics[width=0.48\textwidth]{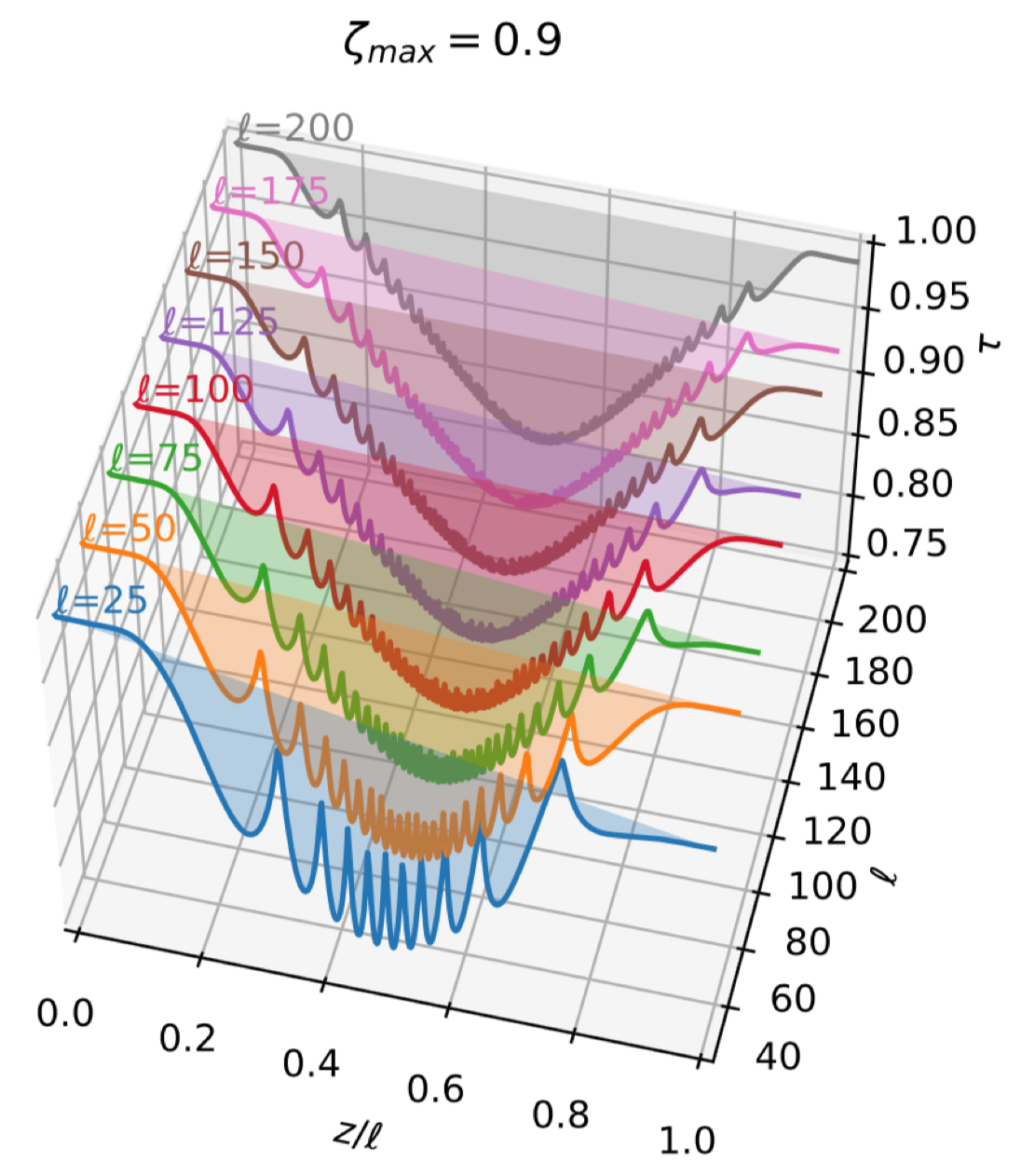}  
\end{center}
\caption{
Intensity plot of the conductance \(\tau\) as a function of \(z/\ell\) and energy \(E\) (in units of \(\hbar v_F / \ell\)). 
The plot reveals the oscillatory behavior of \(\tau\), arising from interference effects during wave propagation through the tilted region. 
This plot corresponds to the bump profile with $\zeta_{\rm max}=0.9$.
}
\label{Fig08}
\end{figure}

Before concluding this section, it is worth addressing why for the bump profile $\zeta_b$ which is symmetric around \( z/\ell = 0.5 \), the solutions are not
symmetric; specifically, why are the solutions at the boundaries \( z/\ell = 0 \) and \( z/\ell = 1 \) not identical, even though layers 1 and 3 are equivalent? 
The asymmetry arises from the reflected waves: while the solutions of the equations in the set \eqref{Eq23} are periodic, there is always a reflected wave, and not all of the incident waves are transmitted. This partial reflection leads to an asymmetrical profile in the solution.  
Figure~\ref{Fig06} illustrates the transmission (\(T\)) and reflection (\(R\)) coefficients, calculated using Eqs.~\eqref{Eq29}, 
for a fixed  \( \ell = 50 \) and different values of \( \zeta_{\text{max}} \), as functions 
of \( \theta_i \) in an interval close to \( \pi/2 \). Furthermore, as pointed out, the spinor $\varphi$ corresponds to a local frame. 
This local frame at every point can be considered as a upright Dirac cone moving with velocity $\zeta v_F$~\cite{Jafari2024}. 
As such, one moving direction is preferred and hence the two directions
along and opposite to moving velocity $v_F\vec\zeta$ are not equivalent, although the profile of {\em magnitude} $\zeta$ looks symmetric.

\section{Transport measurement of red-shift pattern}
\label{Sec05}
To suggest a tangible experimental trace of the above red-shift features, in Fig.~\ref{FigTransmission} we have computed the integrated transmission \(\tau\) 
that quantifies the fraction of the incident wave transmitted through the tilted region by
\[
\tau = \frac{2}{\pi} \int_0^{\pi/2} T(\theta_i) \, d\theta_i,
\]  
where \(T(\theta_i)\) is the angle-dependent transmission coefficient. For this analysis, we evaluate \(\tau\) at various positions \(z/\ell\) within the tilted region, representing the fraction of the length that the wave has traversed.  

Fig.~\ref{FigTransmission} shows \(\tau\) as a function of \(z/\ell\) for the tilting function \(\zeta(z)\). As expected the redshift behavior obtained for a fixed value of the
incident wave parameter $\theta_i$, persists after integrating over $\theta_i$. Again the left-right asymmetry continues to hold in \(\tau\) 
illustrating the complex interplay between wave reflection and transmission within the tilted region. 
These oscillations arise from the interference of waves as they propagate and reflect, and they persist even though the external layers are equivalent.  
The inset of Fig.~\ref{FigTransmission} highlights the corresponding wavenumber, extracted from the oscillatory behavior of \(\tau\). 
Interestingly, the wavenumber matches perfectly with  \(\Delta k_z\) as previously discussed and shown in the inset of Fig.~\ref{Fig05}. 
But here we are dealing with $\theta_i$-integrated version of numerical and analytical results.
This agreement confirms the consistency of the observed oscillatory behavior across different physical quantities, further validating our analytical 
predictions regarding \(\Delta k_z\).  

The above results are obtained for a fixed value of the energy $\ell=200$. For the fixed $E=1$ that also fixes the unit of length $\hbar v_F/E$, it means that the
physical length of the middle layer hosting the variable tilt medium is $200 \hbar v_F/E$. For example when $E=1$ eV and $v_F\sim 10^6$ m/s, $\ell=200$ would correspond to 
$827$ nm. One could ask what happens when the physical junction length becomes thiner. 
This computation has been plotted in Fig.~\ref{Fig08} where the integrated transmission has been plotted for smaller length scales down to nearly an order of magnitude
smaller length scale of $\ell=25$ that at $E=1$ eV corresponds to $10.3$ nm. The following points can be observed in the figures:
\begin{enumerate}
   \item For all the above length scales, the redshift pattern is manifest, i.e. by moving away from the bump, the wave length of the
   oscillations increases. 
   \item For all of the junction lengths, the transmission at the end of junction is around $\sim 0.98$. This should be contrasted 
   with a situation where $\zeta$ abruptly changes across the junction~\cite{AlMarzoog} where as can be seen in Fig. 7 of Ref.~\cite{AlMarzoog},
   for a constant tilt of $\zeta=0.9$ the transmission drops below $0.7$. 
   \item The number of peaks scales linearly with $\ell$ as given in table~\ref{tabpeaks} and fits remarkably well to the formula $n_{\rm p}=0.4\ell$.
   This means that although the wave length undergoes a red-shift pattern away from the bump center, the total number of oscillations for the bump-shaped
   tilting profile manages to scale with the length of the junction. 
   \item As can be clearly seen, by spreading the bump over a longer junction {\em amplitude} of the oscillations steadily decreases for larger $\ell$. 
   One can extrapolate from Fig.~\ref{Fig08} that for very large junction length $\ell$, the transmission curve becomes a smooth curve. 
\end{enumerate}

\begin{table}[t]
\begin{tabular}{|c|cccccccc|}
\hline
$n_{\rm p}$ & 10 & 20& 30& 40& 50& 60& 70& 79\\
\hline
$\ell$      & 25&  50& 75& 100& 125& 150& 175& 200\\
\hline
\end{tabular}
\caption{Number of conduction peaks $n_{\rm p}$ versus the length $\ell$ of the junction in Fig.~\ref{Fig08}.}
\label{tabpeaks}
\end{table}

The property (2) above rests on the fact that gradual change in $\zeta$ that gives rise to gradual change in velocity, produces 
small impedance mismatch~\cite{GeorgiWaves}, thereby diminishing the reflection and enhancing the transmission. This property can be employed as engineering
concept in designing media to convert electron wave number while the nearly perfectly transmit. 
The wave length of the incident and transmitted waves are red-/blue-shifted depending on the step direction of the sigmoid profile. 
This concept is similar to the graded-index optical fiber~\cite{Gong2010,Gong2014}. 

For larger $\ell$, the bump in Fig.~\ref{Fig03} spreads over a longer distance, and therefore the spatial variations of $\zeta$, namely $\partial_z \zeta$
on average becomes smaller. In such a situation, the property (4) above means that the amplitude of the oscillations are controlled by the derivative $\partial_z\zeta$. 
Such derivatives appear in the Christoffel connections that govern the motion of waves in the ray-optics limit. 

In experimental setup measuring the conductance one usually synthesizes one sample with a fixed $\ell$ and there is not much room to change the $\ell$. How can the
above results be useful? As pointed out, the linear scaling of energy and momentum implies that for a physical junction with a given length, the dimensionless
number $\ell$ that counts the number of $\lambda_E=\hbar v_F/E$ fitting into the physical junction, can be decreased by increasing $\lambda_E$. Therefore
moving to lower $E$ corresponds to moving from gray plot to blue plot in Fig.~\ref{Fig08}. This can be done in the lab by sweeping the voltage in 
a scanning tunneling spectroscopy: Scanning along the junction can probe various $z/\ell$ positions, whereas scanning the energy from low to high 
voltages amounts to scanning from blue to gray curves of Fig.~\ref{Fig08}. Measuring the tunneling current at a fixe energy as a function of $z/\ell$ will
give a typical redshift pattern of the form displayed in Fig.~\ref{FigTransmission}. Increasing the voltage and scanning again the junction width would lead to
(i) increase in the number of peaks, and (ii) decrease in the amplitude of oscillations as in Fig.~\ref{Fig08}. For example doubling the voltage and scanning 
the junction width $z/\ell=0$ to $z/\ell=1$ is expected to double the number of peaks within the junction as suggested by table~\ref{tabpeaks}. 

\section{Summary and outlook}
\label{Sec06}
Tilt of Dirac cones is a proxy for an emergent spacetime metric where tilting parameter directly appears in the off-diagonal components of the 
spacetime metric that mix space and time coordinates. 
When an underlying metric is present, it must show up in various quantities. In this work we have computed the transmission of electron waves
through a region where tilting has spatial profiles of bump and sigmoid shape in Fig.~\ref{Fig03}. We find features in the
amplitude of the wave function, spherical angles of the pseudospin, and the transmission coefficients that resemble the "gravitational redshift". 
It means that the wave length of oscillatory behavior is smaller in regions with larger tilt $\zeta$ and get (red) shifted to larger wave lengths
in regions with smaller $\zeta$. 

The ability to red- or blue-shift electron waves, in addition to demonstrating a general-relativistic effect in the lab, can find potential
application: Since due to very low impedance mismatch that arises from gradual change of tilting, the reflection is diminished and the medium
nearly perfectly transmits the waves a sigmoid tilting profile can be used to perfectly transmit electron waves and deliver them with red-/blue-shifted
wave lengths. 

Recent experimental evidence for the tilting induced by magnetic influence in the spin-orbit coupled Dirac cones at the surface of topological
insulators~\cite{Ogawa2016,Jafari2024} is a promising platform to imprint spatially variable textures of tilting ${\boldsymbol{\zeta}}$ by a corresponding
texture of the magnetization. The present proof of principle study indicates that aspects of the underlying spacetime geometry can be detected by
standard solid-state spectroscopic tools. In the spin-orbit coupled Dirac cones, interesting oscillatory behavior of pseudo-spin orientation and 
the corresponding red-shift effects is expected to find novel applications in spintronics devices.

\section{Acknowledgements} 
S.A.J. acknowledges the support of Hendrik J. Bluhm.

\bibliography{Refs}

\begin{thebibliography}{48}%
\makeatletter
\providecommand \@ifxundefined [1]{%
 \@ifx{#1\undefined}
}%
\providecommand \@ifnum [1]{%
 \ifnum #1\expandafter \@firstoftwo
 \else \expandafter \@secondoftwo
 \fi
}%
\providecommand \@ifx [1]{%
 \ifx #1\expandafter \@firstoftwo
 \else \expandafter \@secondoftwo
 \fi
}%
\providecommand \natexlab [1]{#1}%
\providecommand \enquote  [1]{``#1''}%
\providecommand \bibnamefont  [1]{#1}%
\providecommand \bibfnamefont [1]{#1}%
\providecommand \citenamefont [1]{#1}%
\providecommand \href@noop [0]{\@secondoftwo}%
\providecommand \href [0]{\begingroup \@sanitize@url \@href}%
\providecommand \@href[1]{\@@startlink{#1}\@@href}%
\providecommand \@@href[1]{\endgroup#1\@@endlink}%
\providecommand \@sanitize@url [0]{\catcode `\\12\catcode `\$12\catcode
  `\&12\catcode `\#12\catcode `\^12\catcode `\_12\catcode `\%12\relax}%
\providecommand \@@startlink[1]{}%
\providecommand \@@endlink[0]{}%
\providecommand \url  [0]{\begingroup\@sanitize@url \@url }%
\providecommand \@url [1]{\endgroup\@href {#1}{\urlprefix }}%
\providecommand \urlprefix  [0]{URL }%
\providecommand \Eprint [0]{\href }%
\providecommand \doibase [0]{https://doi.org/}%
\providecommand \selectlanguage [0]{\@gobble}%
\providecommand \bibinfo  [0]{\@secondoftwo}%
\providecommand \bibfield  [0]{\@secondoftwo}%
\providecommand \translation [1]{[#1]}%
\providecommand \BibitemOpen [0]{}%
\providecommand \bibitemStop [0]{}%
\providecommand \bibitemNoStop [0]{.\EOS\space}%
\providecommand \EOS [0]{\spacefactor3000\relax}%
\providecommand \BibitemShut  [1]{\csname bibitem#1\endcsname}%
\let\auto@bib@innerbib\@empty
\bibitem [{\citenamefont {Muechler}\ \emph {et~al.}(2016)\citenamefont
  {Muechler}, \citenamefont {Alexandradinata}, \citenamefont {Neupert},\ and\
  \citenamefont {Car}}]{Muechler2016}%
  \BibitemOpen
  \bibfield  {author} {\bibinfo {author} {\bibfnamefont {L.}~\bibnamefont
  {Muechler}}, \bibinfo {author} {\bibfnamefont {A.}~\bibnamefont
  {Alexandradinata}}, \bibinfo {author} {\bibfnamefont {T.}~\bibnamefont
  {Neupert}},\ and\ \bibinfo {author} {\bibfnamefont {R.}~\bibnamefont {Car}},\
  }\bibfield  {title} {\bibinfo {title} {Topological nonsymmorphic metals from
  band inversion},\ }\bibfield  {journal} {\bibinfo  {journal} {Physical Review
  X}\ }\textbf {\bibinfo {volume} {6}},\ \href
  {https://doi.org/10.1103/physrevx.6.041069} {10.1103/physrevx.6.041069}
  (\bibinfo {year} {2016})\BibitemShut {NoStop}%
\bibitem [{\citenamefont {Isobe}(2017)}]{Isobe2017}%
  \BibitemOpen
  \bibfield  {author} {\bibinfo {author} {\bibfnamefont {H.}~\bibnamefont
  {Isobe}},\ }\bibinfo {title} {Tilted dirac cones in two dimensions},\ in\
  \href {https://doi.org/10.1007/978-981-10-3743-6_3} {\emph {\bibinfo
  {booktitle} {Theoretical Study on Correlation Effects in Topological
  Matter}}}\ (\bibinfo  {publisher} {Springer Singapore},\ \bibinfo {year}
  {2017})\ p.\ \bibinfo {pages} {63–81}\BibitemShut {NoStop}%
\bibitem [{\citenamefont {Kobayashi}\ \emph {et~al.}(2007)\citenamefont
  {Kobayashi}, \citenamefont {Katayama}, \citenamefont {Suzumura},\ and\
  \citenamefont {Fukuyama}}]{Kobayashi2007}%
  \BibitemOpen
  \bibfield  {author} {\bibinfo {author} {\bibfnamefont {A.}~\bibnamefont
  {Kobayashi}}, \bibinfo {author} {\bibfnamefont {S.}~\bibnamefont {Katayama}},
  \bibinfo {author} {\bibfnamefont {Y.}~\bibnamefont {Suzumura}},\ and\
  \bibinfo {author} {\bibfnamefont {H.}~\bibnamefont {Fukuyama}},\ }\bibfield
  {title} {\bibinfo {title} {Massless fermions in organic conductor},\ }\href
  {https://doi.org/10.1143/jpsj.76.034711} {\bibfield  {journal} {\bibinfo
  {journal} {Journal of the Physical Society of Japan}\ }\textbf {\bibinfo
  {volume} {76}},\ \bibinfo {pages} {034711} (\bibinfo {year}
  {2007})}\BibitemShut {NoStop}%
\bibitem [{\citenamefont {Varykhalov}\ \emph {et~al.}(2017)\citenamefont
  {Varykhalov}, \citenamefont {Marchenko}, \citenamefont {Sánchez-Barriga},
  \citenamefont {Golias}, \citenamefont {Rader},\ and\ \citenamefont
  {Bihlmayer}}]{Varykhalov2017}%
  \BibitemOpen
  \bibfield  {author} {\bibinfo {author} {\bibfnamefont {A.}~\bibnamefont
  {Varykhalov}}, \bibinfo {author} {\bibfnamefont {D.}~\bibnamefont
  {Marchenko}}, \bibinfo {author} {\bibfnamefont {J.}~\bibnamefont
  {Sánchez-Barriga}}, \bibinfo {author} {\bibfnamefont {E.}~\bibnamefont
  {Golias}}, \bibinfo {author} {\bibfnamefont {O.}~\bibnamefont {Rader}},\ and\
  \bibinfo {author} {\bibfnamefont {G.}~\bibnamefont {Bihlmayer}},\ }\bibfield
  {title} {\bibinfo {title} {Tilted dirac cone on w(110) protected by mirror
  symmetry},\ }\bibfield  {journal} {\bibinfo  {journal} {Physical Review B}\
  }\textbf {\bibinfo {volume} {95}},\ \href
  {https://doi.org/10.1103/physrevb.95.245421} {10.1103/physrevb.95.245421}
  (\bibinfo {year} {2017})\BibitemShut {NoStop}%
\bibitem [{\citenamefont {Lopez-Bezanilla}\ and\ \citenamefont
  {Littlewood}(2016)}]{LopezBezanilla2016}%
  \BibitemOpen
  \bibfield  {author} {\bibinfo {author} {\bibfnamefont {A.}~\bibnamefont
  {Lopez-Bezanilla}}\ and\ \bibinfo {author} {\bibfnamefont {P.~B.}\
  \bibnamefont {Littlewood}},\ }\bibfield  {title} {\bibinfo {title}
  {Electronic properties of $8pmmn$ borophene},\ }\bibfield  {journal}
  {\bibinfo  {journal} {Physical Review B}\ }\textbf {\bibinfo {volume} {93}},\
  \href {https://doi.org/10.1103/physrevb.93.241405}
  {10.1103/physrevb.93.241405} (\bibinfo {year} {2016})\BibitemShut {NoStop}%
\bibitem [{\citenamefont {Zhang}\ \emph {et~al.}(2023)\citenamefont {Zhang},
  \citenamefont {Shao}, \citenamefont {Wang}, \citenamefont {Yang},
  \citenamefont {Yang},\ and\ \citenamefont {Tsymbal}}]{Zhang2023}%
  \BibitemOpen
  \bibfield  {author} {\bibinfo {author} {\bibfnamefont {S.-H.}\ \bibnamefont
  {Zhang}}, \bibinfo {author} {\bibfnamefont {D.-F.}\ \bibnamefont {Shao}},
  \bibinfo {author} {\bibfnamefont {Z.-A.}\ \bibnamefont {Wang}}, \bibinfo
  {author} {\bibfnamefont {J.}~\bibnamefont {Yang}}, \bibinfo {author}
  {\bibfnamefont {W.}~\bibnamefont {Yang}},\ and\ \bibinfo {author}
  {\bibfnamefont {E.~Y.}\ \bibnamefont {Tsymbal}},\ }\bibfield  {title}
  {\bibinfo {title} {Tunneling valley hall effect driven by tilted dirac
  fermions},\ }\bibfield  {journal} {\bibinfo  {journal} {Physical Review
  Letters}\ }\textbf {\bibinfo {volume} {131}},\ \href
  {https://doi.org/10.1103/physrevlett.131.246301}
  {10.1103/physrevlett.131.246301} (\bibinfo {year} {2023})\BibitemShut
  {NoStop}%
\bibitem [{Tan()}]{Tan2021}%
  \BibitemOpen
  \bibfield  {title} {\bibinfo {title} {Anisotropic longitudinal optical
  conductivities of tilted dirac bands in $1t'-${M}o{S}$_2$},\ }\href@noop {}
  {\ }\BibitemShut {NoStop}%
\bibitem [{\citenamefont {Park}\ \emph {et~al.}(2022)\citenamefont {Park},
  \citenamefont {Sammon}, \citenamefont {Mele},\ and\ \citenamefont
  {Low}}]{Park2022}%
  \BibitemOpen
  \bibfield  {author} {\bibinfo {author} {\bibfnamefont {S.~H.}\ \bibnamefont
  {Park}}, \bibinfo {author} {\bibfnamefont {M.}~\bibnamefont {Sammon}},
  \bibinfo {author} {\bibfnamefont {E.}~\bibnamefont {Mele}},\ and\ \bibinfo
  {author} {\bibfnamefont {T.}~\bibnamefont {Low}},\ }\bibfield  {title}
  {\bibinfo {title} {Plasmonic gain in current biased tilted dirac nodes},\
  }\bibfield  {journal} {\bibinfo  {journal} {Nature Communications}\ }\textbf
  {\bibinfo {volume} {13}},\ \href {https://doi.org/10.1038/s41467-022-35139-y}
  {10.1038/s41467-022-35139-y} (\bibinfo {year} {2022})\BibitemShut {NoStop}%
\bibitem [{\citenamefont {Jalali-Mola}\ and\ \citenamefont
  {Jafari}(2018{\natexlab{a}})}]{JalaliMola2018}%
  \BibitemOpen
  \bibfield  {author} {\bibinfo {author} {\bibfnamefont {Z.}~\bibnamefont
  {Jalali-Mola}}\ and\ \bibinfo {author} {\bibfnamefont {S.~A.}\ \bibnamefont
  {Jafari}},\ }\bibfield  {title} {\bibinfo {title} {Tilt-induced kink in the
  plasmon dispersion of two-dimensional dirac electrons},\ }\bibfield
  {journal} {\bibinfo  {journal} {Physical Review B}\ }\textbf {\bibinfo
  {volume} {98}},\ \href {https://doi.org/10.1103/physrevb.98.195415}
  {10.1103/physrevb.98.195415} (\bibinfo {year}
  {2018}{\natexlab{a}})\BibitemShut {NoStop}%
\bibitem [{\citenamefont {Jalali-Mola}\ and\ \citenamefont
  {Jafari}(2018{\natexlab{b}})}]{JalaliMola2018double}%
  \BibitemOpen
  \bibfield  {author} {\bibinfo {author} {\bibfnamefont {Z.}~\bibnamefont
  {Jalali-Mola}}\ and\ \bibinfo {author} {\bibfnamefont {S.~A.}\ \bibnamefont
  {Jafari}},\ }\bibfield  {title} {\bibinfo {title} {Kinked plasmon dispersion
  in borophene-borophene and borophene-graphene double layers},\ }\bibfield
  {journal} {\bibinfo  {journal} {Physical Review B}\ }\textbf {\bibinfo
  {volume} {98}},\ \href {https://doi.org/10.1103/physrevb.98.235430}
  {10.1103/physrevb.98.235430} (\bibinfo {year}
  {2018}{\natexlab{b}})\BibitemShut {NoStop}%
\bibitem [{\citenamefont {Torbatian}\ \emph {et~al.}(2021)\citenamefont
  {Torbatian}, \citenamefont {Novko},\ and\ \citenamefont
  {Asgari}}]{Torbatian2021}%
  \BibitemOpen
  \bibfield  {author} {\bibinfo {author} {\bibfnamefont {Z.}~\bibnamefont
  {Torbatian}}, \bibinfo {author} {\bibfnamefont {D.}~\bibnamefont {Novko}},\
  and\ \bibinfo {author} {\bibfnamefont {R.}~\bibnamefont {Asgari}},\
  }\bibfield  {title} {\bibinfo {title} {Hyperbolic plasmon modes in tilted
  dirac cone phases of borophene},\ }\bibfield  {journal} {\bibinfo  {journal}
  {Physical Review B}\ }\textbf {\bibinfo {volume} {104}},\ \href
  {https://doi.org/10.1103/physrevb.104.075432} {10.1103/physrevb.104.075432}
  (\bibinfo {year} {2021})\BibitemShut {NoStop}%
\bibitem [{\citenamefont {Mojarro}\ \emph {et~al.}(2022)\citenamefont
  {Mojarro}, \citenamefont {Carrillo-Bastos},\ and\ \citenamefont
  {Maytorena}}]{Mojarro2022}%
  \BibitemOpen
  \bibfield  {author} {\bibinfo {author} {\bibfnamefont {M.~A.}\ \bibnamefont
  {Mojarro}}, \bibinfo {author} {\bibfnamefont {R.}~\bibnamefont
  {Carrillo-Bastos}},\ and\ \bibinfo {author} {\bibfnamefont {J.~A.}\
  \bibnamefont {Maytorena}},\ }\bibfield  {title} {\bibinfo {title} {Hyperbolic
  plasmons in massive tilted two-dimensional dirac materials},\ }\bibfield
  {journal} {\bibinfo  {journal} {Physical Review B}\ }\textbf {\bibinfo
  {volume} {105}},\ \href {https://doi.org/10.1103/physrevb.105.l201408}
  {10.1103/physrevb.105.l201408} (\bibinfo {year} {2022})\BibitemShut {NoStop}%
\bibitem [{\citenamefont {Sengupta}\ \emph {et~al.}(2018)\citenamefont
  {Sengupta}, \citenamefont {Tan}, \citenamefont {Bellotti},\ and\
  \citenamefont {Shi}}]{Sengupta2018}%
  \BibitemOpen
  \bibfield  {author} {\bibinfo {author} {\bibfnamefont {P.}~\bibnamefont
  {Sengupta}}, \bibinfo {author} {\bibfnamefont {Y.}~\bibnamefont {Tan}},
  \bibinfo {author} {\bibfnamefont {E.}~\bibnamefont {Bellotti}},\ and\
  \bibinfo {author} {\bibfnamefont {J.}~\bibnamefont {Shi}},\ }\bibfield
  {title} {\bibinfo {title} {Anomalous heat flow in 8-pmmn borophene with
  tilted dirac cones},\ }\href {https://doi.org/10.1088/1361-648x/aae111}
  {\bibfield  {journal} {\bibinfo  {journal} {Journal of Physics: Condensed
  Matter}\ }\textbf {\bibinfo {volume} {30}},\ \bibinfo {pages} {435701}
  (\bibinfo {year} {2018})}\BibitemShut {NoStop}%
\bibitem [{\citenamefont {KAWARABAYASHI}\ \emph {et~al.}(2012)\citenamefont
  {KAWARABAYASHI}, \citenamefont {HATSUGAI}, \citenamefont {MORIMOTO},\ and\
  \citenamefont {AOKI}}]{KAWARABAYASHI2012}%
  \BibitemOpen
  \bibfield  {author} {\bibinfo {author} {\bibfnamefont {T.}~\bibnamefont
  {KAWARABAYASHI}}, \bibinfo {author} {\bibfnamefont {Y.}~\bibnamefont
  {HATSUGAI}}, \bibinfo {author} {\bibfnamefont {T.}~\bibnamefont {MORIMOTO}},\
  and\ \bibinfo {author} {\bibfnamefont {H.}~\bibnamefont {AOKI}},\ }\bibfield
  {title} {\bibinfo {title} {Generalization of chiral symmetry for tilted dirac
  cones},\ }\href {https://doi.org/10.1142/s2010194512006046} {\bibfield
  {journal} {\bibinfo  {journal} {International Journal of Modern Physics:
  Conference Series}\ }\textbf {\bibinfo {volume} {11}},\ \bibinfo {pages}
  {145–150} (\bibinfo {year} {2012})}\BibitemShut {NoStop}%
\bibitem [{\citenamefont {Ohki}\ \emph {et~al.}(2020)\citenamefont {Ohki},
  \citenamefont {Hirata}, \citenamefont {Tani}, \citenamefont {Kanoda},\ and\
  \citenamefont {Kobayashi}}]{Ohki2020}%
  \BibitemOpen
  \bibfield  {author} {\bibinfo {author} {\bibfnamefont {D.}~\bibnamefont
  {Ohki}}, \bibinfo {author} {\bibfnamefont {M.}~\bibnamefont {Hirata}},
  \bibinfo {author} {\bibfnamefont {T.}~\bibnamefont {Tani}}, \bibinfo {author}
  {\bibfnamefont {K.}~\bibnamefont {Kanoda}},\ and\ \bibinfo {author}
  {\bibfnamefont {A.}~\bibnamefont {Kobayashi}},\ }\bibfield  {title} {\bibinfo
  {title} {Chiral excitonic instability of two-dimensional tilted dirac
  cones},\ }\bibfield  {journal} {\bibinfo  {journal} {Physical Review
  Research}\ }\textbf {\bibinfo {volume} {2}},\ \href
  {https://doi.org/10.1103/physrevresearch.2.033479}
  {10.1103/physrevresearch.2.033479} (\bibinfo {year} {2020})\BibitemShut
  {NoStop}%
\bibitem [{\citenamefont {Gomes}\ and\ \citenamefont
  {Ramos}(2021)}]{Gomes2021}%
  \BibitemOpen
  \bibfield  {author} {\bibinfo {author} {\bibfnamefont {Y.~M.~P.}\
  \bibnamefont {Gomes}}\ and\ \bibinfo {author} {\bibfnamefont {R.~O.}\
  \bibnamefont {Ramos}},\ }\bibfield  {title} {\bibinfo {title} {Tilted dirac
  cone effects and chiral symmetry breaking in a planar four-fermion model},\
  }\bibfield  {journal} {\bibinfo  {journal} {Physical Review B}\ }\textbf
  {\bibinfo {volume} {104}},\ \href
  {https://doi.org/10.1103/physrevb.104.245111} {10.1103/physrevb.104.245111}
  (\bibinfo {year} {2021})\BibitemShut {NoStop}%
\bibitem [{\citenamefont {Zhao}\ and\ \citenamefont {Wang}(2019)}]{Zhao2019}%
  \BibitemOpen
  \bibfield  {author} {\bibinfo {author} {\bibfnamefont {P.-L.}\ \bibnamefont
  {Zhao}}\ and\ \bibinfo {author} {\bibfnamefont {A.-M.}\ \bibnamefont
  {Wang}},\ }\bibfield  {title} {\bibinfo {title} {Interplay between tilt,
  disorder, and coulomb interaction in type-i dirac fermions},\ }\bibfield
  {journal} {\bibinfo  {journal} {Physical Review B}\ }\textbf {\bibinfo
  {volume} {100}},\ \href {https://doi.org/10.1103/physrevb.100.125138}
  {10.1103/physrevb.100.125138} (\bibinfo {year} {2019})\BibitemShut {NoStop}%
\bibitem [{\citenamefont {Mojarro}\ \emph {et~al.}(2023)\citenamefont
  {Mojarro}, \citenamefont {Carrillo-Bastos},\ and\ \citenamefont
  {Maytorena}}]{Mojarro2023}%
  \BibitemOpen
  \bibfield  {author} {\bibinfo {author} {\bibfnamefont {M.~A.}\ \bibnamefont
  {Mojarro}}, \bibinfo {author} {\bibfnamefont {R.}~\bibnamefont
  {Carrillo-Bastos}},\ and\ \bibinfo {author} {\bibfnamefont {J.~A.}\
  \bibnamefont {Maytorena}},\ }\bibfield  {title} {\bibinfo {title} {Thermal
  difference reflectivity of tilted two-dimensional dirac materials},\
  }\bibfield  {journal} {\bibinfo  {journal} {Physical Review B}\ }\textbf
  {\bibinfo {volume} {108}},\ \href
  {https://doi.org/10.1103/physrevb.108.l161401} {10.1103/physrevb.108.l161401}
  (\bibinfo {year} {2023})\BibitemShut {NoStop}%
\bibitem [{\citenamefont {Paul}\ \emph {et~al.}(2019)\citenamefont {Paul},
  \citenamefont {Islam},\ and\ \citenamefont {Saha}}]{Paul2019}%
  \BibitemOpen
  \bibfield  {author} {\bibinfo {author} {\bibfnamefont {G.~C.}\ \bibnamefont
  {Paul}}, \bibinfo {author} {\bibfnamefont {S.~F.}\ \bibnamefont {Islam}},\
  and\ \bibinfo {author} {\bibfnamefont {A.}~\bibnamefont {Saha}},\ }\bibfield
  {title} {\bibinfo {title} {Fingerprints of tilted dirac cones on the rkky
  exchange interaction in 8- pmmn borophene},\ }\bibfield  {journal} {\bibinfo
  {journal} {Physical Review B}\ }\textbf {\bibinfo {volume} {99}},\ \href
  {https://doi.org/10.1103/physrevb.99.155418} {10.1103/physrevb.99.155418}
  (\bibinfo {year} {2019})\BibitemShut {NoStop}%
\bibitem [{\citenamefont {Sinha}(2019)}]{Sinha2019}%
  \BibitemOpen
  \bibfield  {author} {\bibinfo {author} {\bibfnamefont {D.}~\bibnamefont
  {Sinha}},\ }\bibfield  {title} {\bibinfo {title} {Spin transport and spin
  pump in graphene-like materials: effects of tilted dirac cone},\ }\bibfield
  {journal} {\bibinfo  {journal} {The European Physical Journal B}\ }\textbf
  {\bibinfo {volume} {92}},\ \href {https://doi.org/10.1140/epjb/e2019-90332-7}
  {10.1140/epjb/e2019-90332-7} (\bibinfo {year} {2019})\BibitemShut {NoStop}%
\bibitem [{\citenamefont {Suzumura}\ \emph {et~al.}(2014)\citenamefont
  {Suzumura}, \citenamefont {Proskurin},\ and\ \citenamefont
  {Ogata}}]{Suzumura2014}%
  \BibitemOpen
  \bibfield  {author} {\bibinfo {author} {\bibfnamefont {Y.}~\bibnamefont
  {Suzumura}}, \bibinfo {author} {\bibfnamefont {I.}~\bibnamefont
  {Proskurin}},\ and\ \bibinfo {author} {\bibfnamefont {M.}~\bibnamefont
  {Ogata}},\ }\bibfield  {title} {\bibinfo {title} {Effect of tilting on the
  in-plane conductivity of dirac electrons in organic conductor},\ }\href
  {https://doi.org/10.7566/jpsj.83.023701} {\bibfield  {journal} {\bibinfo
  {journal} {Journal of the Physical Society of Japan}\ }\textbf {\bibinfo
  {volume} {83}},\ \bibinfo {pages} {023701} (\bibinfo {year}
  {2014})}\BibitemShut {NoStop}%
\bibitem [{\citenamefont {Volovik}\ and\ \citenamefont
  {Zhang}(2017)}]{Volovik2017}%
  \BibitemOpen
  \bibfield  {author} {\bibinfo {author} {\bibfnamefont {G.~E.}\ \bibnamefont
  {Volovik}}\ and\ \bibinfo {author} {\bibfnamefont {K.}~\bibnamefont
  {Zhang}},\ }\bibfield  {title} {\bibinfo {title} {Lifshitz transitions,
  type-ii dirac and weyl fermions, event horizon and all that},\ }\href
  {https://doi.org/10.1007/s10909-017-1817-8} {\bibfield  {journal} {\bibinfo
  {journal} {Journal of Low Temperature Physics}\ }\textbf {\bibinfo {volume}
  {189}},\ \bibinfo {pages} {276–299} (\bibinfo {year} {2017})}\BibitemShut
  {NoStop}%
\bibitem [{\citenamefont {Volovik}(2021)}]{Volovik2021}%
  \BibitemOpen
  \bibfield  {author} {\bibinfo {author} {\bibfnamefont {G.~E.}\ \bibnamefont
  {Volovik}},\ }\bibfield  {title} {\bibinfo {title} {Type-ii weyl semimetal
  versus gravastar},\ }\href {https://doi.org/10.1134/s0021364021160013}
  {\bibfield  {journal} {\bibinfo  {journal} {JETP Letters}\ }\textbf {\bibinfo
  {volume} {114}},\ \bibinfo {pages} {236–242} (\bibinfo {year}
  {2021})}\BibitemShut {NoStop}%
\bibitem [{\citenamefont {Liang}\ and\ \citenamefont
  {Ojanen}(2019)}]{Liang2019}%
  \BibitemOpen
  \bibfield  {author} {\bibinfo {author} {\bibfnamefont {L.}~\bibnamefont
  {Liang}}\ and\ \bibinfo {author} {\bibfnamefont {T.}~\bibnamefont {Ojanen}},\
  }\bibfield  {title} {\bibinfo {title} {Curved spacetime theory of
  inhomogeneous weyl materials},\ }\bibfield  {journal} {\bibinfo  {journal}
  {Physical Review Research}\ }\textbf {\bibinfo {volume} {1}},\ \href
  {https://doi.org/10.1103/physrevresearch.1.032006}
  {10.1103/physrevresearch.1.032006} (\bibinfo {year} {2019})\BibitemShut
  {NoStop}%
\bibitem [{\citenamefont {Zubkov}(2018)}]{Zubkov2018}%
  \BibitemOpen
  \bibfield  {author} {\bibinfo {author} {\bibfnamefont {M.}~\bibnamefont
  {Zubkov}},\ }\bibfield  {title} {\bibinfo {title} {Analogies between the
  black hole interior and the type ii weyl semimetals},\ }\href
  {https://doi.org/10.3390/universe4120135} {\bibfield  {journal} {\bibinfo
  {journal} {Universe}\ }\textbf {\bibinfo {volume} {4}},\ \bibinfo {pages}
  {135} (\bibinfo {year} {2018})}\BibitemShut {NoStop}%
\bibitem [{\citenamefont {Volovik}(2016)}]{Volovik2016}%
  \BibitemOpen
  \bibfield  {author} {\bibinfo {author} {\bibfnamefont {G.~E.}\ \bibnamefont
  {Volovik}},\ }\bibfield  {title} {\bibinfo {title} {Black hole and hawking
  radiation by type-ii weyl fermions},\ }\href
  {https://doi.org/10.1134/s0021364016210050} {\bibfield  {journal} {\bibinfo
  {journal} {JETP Letters}\ }\textbf {\bibinfo {volume} {104}},\ \bibinfo
  {pages} {645–648} (\bibinfo {year} {2016})}\BibitemShut {NoStop}%
\bibitem [{\citenamefont {Kedem}\ \emph {et~al.}(2020)\citenamefont {Kedem},
  \citenamefont {Bergholtz},\ and\ \citenamefont {Wilczek}}]{Kedem2020}%
  \BibitemOpen
  \bibfield  {author} {\bibinfo {author} {\bibfnamefont {Y.}~\bibnamefont
  {Kedem}}, \bibinfo {author} {\bibfnamefont {E.~J.}\ \bibnamefont
  {Bergholtz}},\ and\ \bibinfo {author} {\bibfnamefont {F.}~\bibnamefont
  {Wilczek}},\ }\bibfield  {title} {\bibinfo {title} {Black and white holes at
  material junctions},\ }\bibfield  {journal} {\bibinfo  {journal} {Physical
  Review Research}\ }\textbf {\bibinfo {volume} {2}},\ \href
  {https://doi.org/10.1103/physrevresearch.2.043285}
  {10.1103/physrevresearch.2.043285} (\bibinfo {year} {2020})\BibitemShut
  {NoStop}%
\bibitem [{\citenamefont {K\"{o}nye}\ \emph {et~al.}(2022)\citenamefont
  {K\"{o}nye}, \citenamefont {Morice}, \citenamefont {Chernyavsky},
  \citenamefont {Moghaddam}, \citenamefont {van~den Brink},\ and\ \citenamefont
  {van Wezel}}]{Konye2022}%
  \BibitemOpen
  \bibfield  {author} {\bibinfo {author} {\bibfnamefont {V.}~\bibnamefont
  {K\"{o}nye}}, \bibinfo {author} {\bibfnamefont {C.}~\bibnamefont {Morice}},
  \bibinfo {author} {\bibfnamefont {D.}~\bibnamefont {Chernyavsky}}, \bibinfo
  {author} {\bibfnamefont {A.~G.}\ \bibnamefont {Moghaddam}}, \bibinfo {author}
  {\bibfnamefont {J.}~\bibnamefont {van~den Brink}},\ and\ \bibinfo {author}
  {\bibfnamefont {J.}~\bibnamefont {van Wezel}},\ }\bibfield  {title} {\bibinfo
  {title} {Horizon physics of quasi-one-dimensional tilted weyl cones on a
  lattice},\ }\bibfield  {journal} {\bibinfo  {journal} {Physical Review
  Research}\ }\textbf {\bibinfo {volume} {4}},\ \href
  {https://doi.org/10.1103/physrevresearch.4.033237}
  {10.1103/physrevresearch.4.033237} (\bibinfo {year} {2022})\BibitemShut
  {NoStop}%
\bibitem [{\citenamefont {Farajollahpour}\ and\ \citenamefont
  {Jafari}(2020)}]{Farajollahpour2020}%
  \BibitemOpen
  \bibfield  {author} {\bibinfo {author} {\bibfnamefont {T.}~\bibnamefont
  {Farajollahpour}}\ and\ \bibinfo {author} {\bibfnamefont {S.~A.}\
  \bibnamefont {Jafari}},\ }\bibfield  {title} {\bibinfo {title} {Synthetic
  non-abelian gauge fields and gravitomagnetic effects in tilted dirac cone
  systems},\ }\bibfield  {journal} {\bibinfo  {journal} {Physical Review
  Research}\ }\textbf {\bibinfo {volume} {2}},\ \href
  {https://doi.org/10.1103/physrevresearch.2.023410}
  {10.1103/physrevresearch.2.023410} (\bibinfo {year} {2020})\BibitemShut
  {NoStop}%
\bibitem [{\citenamefont {Hashimoto}\ and\ \citenamefont
  {Matsuo}(2020)}]{Hashimoto2020}%
  \BibitemOpen
  \bibfield  {author} {\bibinfo {author} {\bibfnamefont {K.}~\bibnamefont
  {Hashimoto}}\ and\ \bibinfo {author} {\bibfnamefont {Y.}~\bibnamefont
  {Matsuo}},\ }\bibfield  {title} {\bibinfo {title} {Escape from black hole
  analogs in materials: Type-ii weyl semimetals and generic edge states},\
  }\bibfield  {journal} {\bibinfo  {journal} {Physical Review B}\ }\textbf
  {\bibinfo {volume} {102}},\ \href
  {https://doi.org/10.1103/physrevb.102.195128} {10.1103/physrevb.102.195128}
  (\bibinfo {year} {2020})\BibitemShut {NoStop}%
\bibitem [{\citenamefont {Yekta}\ \emph {et~al.}(2023)\citenamefont {Yekta},
  \citenamefont {Hadipour},\ and\ \citenamefont {Jafari}}]{Yekta2023}%
  \BibitemOpen
  \bibfield  {author} {\bibinfo {author} {\bibfnamefont {Y.}~\bibnamefont
  {Yekta}}, \bibinfo {author} {\bibfnamefont {H.}~\bibnamefont {Hadipour}},\
  and\ \bibinfo {author} {\bibfnamefont {S.~A.}\ \bibnamefont {Jafari}},\
  }\bibfield  {title} {\bibinfo {title} {Tunning the tilt of the dirac cone by
  atomic manipulations in 8pmmn borophene},\ }\bibfield  {journal} {\bibinfo
  {journal} {Communications Physics}\ }\textbf {\bibinfo {volume} {6}},\ \href
  {https://doi.org/10.1038/s42005-023-01161-9} {10.1038/s42005-023-01161-9}
  (\bibinfo {year} {2023})\BibitemShut {NoStop}%
\bibitem [{\citenamefont {Ghorashi}\ \emph {et~al.}(2020)\citenamefont
  {Ghorashi}, \citenamefont {Karcher}, \citenamefont {Davis},\ and\
  \citenamefont {Foster}}]{Ghorashi2020}%
  \BibitemOpen
  \bibfield  {author} {\bibinfo {author} {\bibfnamefont {S.~A.~A.}\
  \bibnamefont {Ghorashi}}, \bibinfo {author} {\bibfnamefont {J.~F.}\
  \bibnamefont {Karcher}}, \bibinfo {author} {\bibfnamefont {S.~M.}\
  \bibnamefont {Davis}},\ and\ \bibinfo {author} {\bibfnamefont {M.~S.}\
  \bibnamefont {Foster}},\ }\bibfield  {title} {\bibinfo {title} {Criticality
  across the energy spectrum from random artificial gravitational lensing in
  two-dimensional dirac superconductors},\ }\bibfield  {journal} {\bibinfo
  {journal} {Physical Review B}\ }\textbf {\bibinfo {volume} {101}},\ \href
  {https://doi.org/10.1103/physrevb.101.214521} {10.1103/physrevb.101.214521}
  (\bibinfo {year} {2020})\BibitemShut {NoStop}%
\bibitem [{\citenamefont {Ogawa}\ \emph {et~al.}(2016)\citenamefont {Ogawa},
  \citenamefont {Yoshimi}, \citenamefont {Yasuda}, \citenamefont {Tsukazaki},
  \citenamefont {Kawasaki},\ and\ \citenamefont {Tokura}}]{Ogawa2016}%
  \BibitemOpen
  \bibfield  {author} {\bibinfo {author} {\bibfnamefont {N.}~\bibnamefont
  {Ogawa}}, \bibinfo {author} {\bibfnamefont {R.}~\bibnamefont {Yoshimi}},
  \bibinfo {author} {\bibfnamefont {K.}~\bibnamefont {Yasuda}}, \bibinfo
  {author} {\bibfnamefont {A.}~\bibnamefont {Tsukazaki}}, \bibinfo {author}
  {\bibfnamefont {M.}~\bibnamefont {Kawasaki}},\ and\ \bibinfo {author}
  {\bibfnamefont {Y.}~\bibnamefont {Tokura}},\ }\bibfield  {title} {\bibinfo
  {title} {Zero-bias photocurrent in ferromagnetic topological insulator},\
  }\bibfield  {journal} {\bibinfo  {journal} {Nature Communications}\ }\textbf
  {\bibinfo {volume} {7}},\ \href {https://doi.org/10.1038/ncomms12246}
  {10.1038/ncomms12246} (\bibinfo {year} {2016})\BibitemShut {NoStop}%
\bibitem [{\citenamefont {Jafari}(2024)}]{Jafari2024}%
  \BibitemOpen
  \bibfield  {author} {\bibinfo {author} {\bibfnamefont {S.~A.}\ \bibnamefont
  {Jafari}},\ }\bibfield  {title} {\bibinfo {title} {Moving frame theory of
  zero-bias photocurrent on the surface of topological insulators},\ }\bibfield
   {journal} {\bibinfo  {journal} {Physical Review Research}\ }\textbf
  {\bibinfo {volume} {6}},\ \href
  {https://doi.org/10.1103/physrevresearch.6.033006}
  {10.1103/physrevresearch.6.033006} (\bibinfo {year} {2024})\BibitemShut
  {NoStop}%
\bibitem [{\citenamefont {Einstein}(1916)}]{Einstein1916}%
  \BibitemOpen
  \bibfield  {author} {\bibinfo {author} {\bibfnamefont {A.}~\bibnamefont
  {Einstein}},\ }\bibfield  {title} {\bibinfo {title} {Die grundlage der
  allgemeinen relativit\"{a}tstheorie},\ }\href
  {https://doi.org/10.1002/andp.19163540702} {\bibfield  {journal} {\bibinfo
  {journal} {Annalen der Physik}\ }\textbf {\bibinfo {volume} {354}},\ \bibinfo
  {pages} {769–822} (\bibinfo {year} {1916})}\BibitemShut {NoStop}%
\bibitem [{\citenamefont {Pound}\ and\ \citenamefont
  {Rebka}(1959)}]{Pound1959}%
  \BibitemOpen
  \bibfield  {author} {\bibinfo {author} {\bibfnamefont {R.~V.}\ \bibnamefont
  {Pound}}\ and\ \bibinfo {author} {\bibfnamefont {G.~A.}\ \bibnamefont
  {Rebka}},\ }\bibfield  {title} {\bibinfo {title} {Gravitational red-shift in
  nuclear resonance},\ }\href {https://doi.org/10.1103/physrevlett.3.439}
  {\bibfield  {journal} {\bibinfo  {journal} {Physical Review Letters}\
  }\textbf {\bibinfo {volume} {3}},\ \bibinfo {pages} {439–441} (\bibinfo
  {year} {1959})}\BibitemShut {NoStop}%
\bibitem [{\citenamefont {Ryder}(2009)}]{Ryder2009}%
  \BibitemOpen
  \bibfield  {author} {\bibinfo {author} {\bibfnamefont {L.}~\bibnamefont
  {Ryder}},\ }\href {https://doi.org/10.1017/cbo9780511809033} {\emph {\bibinfo
  {title} {Introduction to General Relativity}}}\ (\bibinfo  {publisher}
  {Cambridge University Press},\ \bibinfo {year} {2009})\BibitemShut {NoStop}%
\bibitem [{\citenamefont {Ko\c{c}}\ and\ \citenamefont
  {T\"{u}t\"{u}nc\"{u}ler}(2003)}]{Koc2003}%
  \BibitemOpen
  \bibfield  {author} {\bibinfo {author} {\bibfnamefont {R.}~\bibnamefont
  {Ko\c{c}}}\ and\ \bibinfo {author} {\bibfnamefont {H.}~\bibnamefont
  {T\"{u}t\"{u}nc\"{u}ler}},\ }\bibfield  {title} {\bibinfo {title} {Exact
  solution of position dependent mass schr\"{o}dinger equation by
  supersymmetric quantum mechanics},\ }\href
  {https://doi.org/10.1002/andp.200351511-1202} {\bibfield  {journal} {\bibinfo
   {journal} {Annalen der Physik}\ }\textbf {\bibinfo {volume} {515}},\
  \bibinfo {pages} {684–691} (\bibinfo {year} {2003})}\BibitemShut {NoStop}%
\bibitem [{\citenamefont {Mazharimousavi}\ and\ \citenamefont
  {Halilsoy}(2013)}]{Mazhari2013}%
  \BibitemOpen
  \bibfield  {author} {\bibinfo {author} {\bibfnamefont {S.~H.}\ \bibnamefont
  {Mazharimousavi}}\ and\ \bibinfo {author} {\bibfnamefont {M.}~\bibnamefont
  {Halilsoy}},\ }\href@noop {} {\bibinfo {title} {One dimensional newton's
  equation with variable mass}} (\bibinfo {year} {2013}),\ \Eprint
  {https://arxiv.org/abs/arXiv:1308.2981} {arXiv:1308.2981} \BibitemShut
  {NoStop}%
\bibitem [{\citenamefont {Rosas-Ortiz}(2020)}]{RosasOrtiz2020}%
  \BibitemOpen
  \bibfield  {author} {\bibinfo {author} {\bibfnamefont {O.}~\bibnamefont
  {Rosas-Ortiz}},\ }\bibinfo {title} {Position-dependent mass systems:
  Classical and quantum pictures},\ in\ \href
  {https://doi.org/10.1007/978-3-030-53305-2_24} {\emph {\bibinfo {booktitle}
  {Geometric Methods in Physics XXXVIII}}}\ (\bibinfo  {publisher} {Springer
  International Publishing},\ \bibinfo {year} {2020})\ p.\ \bibinfo {pages}
  {351–361}\BibitemShut {NoStop}%
\bibitem [{\citenamefont {Raoux}\ \emph {et~al.}(2010)\citenamefont {Raoux},
  \citenamefont {Polini}, \citenamefont {Asgari}, \citenamefont {Hamilton},
  \citenamefont {Fazio},\ and\ \citenamefont {MacDonald}}]{Fazio}%
  \BibitemOpen
  \bibfield  {author} {\bibinfo {author} {\bibfnamefont {A.}~\bibnamefont
  {Raoux}}, \bibinfo {author} {\bibfnamefont {M.}~\bibnamefont {Polini}},
  \bibinfo {author} {\bibfnamefont {R.}~\bibnamefont {Asgari}}, \bibinfo
  {author} {\bibfnamefont {A.~R.}\ \bibnamefont {Hamilton}}, \bibinfo {author}
  {\bibfnamefont {R.}~\bibnamefont {Fazio}},\ and\ \bibinfo {author}
  {\bibfnamefont {A.~H.}\ \bibnamefont {MacDonald}},\ }\bibfield  {title}
  {\bibinfo {title} {Velocity-modulation control of electron-wave propagation
  in graphene},\ }\href {https://doi.org/10.1103/PhysRevB.81.073407} {\bibfield
   {journal} {\bibinfo  {journal} {Phys. Rev. B}\ }\textbf {\bibinfo {volume}
  {81}},\ \bibinfo {pages} {073407} (\bibinfo {year} {2010})}\BibitemShut
  {NoStop}%
\bibitem [{\citenamefont {Joy}\ \emph {et~al.}(2020)\citenamefont {Joy},
  \citenamefont {Khalid},\ and\ \citenamefont {Skinner}}]{Joy}%
  \BibitemOpen
  \bibfield  {author} {\bibinfo {author} {\bibfnamefont {S.}~\bibnamefont
  {Joy}}, \bibinfo {author} {\bibfnamefont {S.}~\bibnamefont {Khalid}},\ and\
  \bibinfo {author} {\bibfnamefont {B.}~\bibnamefont {Skinner}},\ }\bibfield
  {title} {\bibinfo {title} {Transparent mirror effect in
  twist-angle-disordered bilayer graphene},\ }\href
  {https://doi.org/10.1103/PhysRevResearch.2.043416} {\bibfield  {journal}
  {\bibinfo  {journal} {Phys. Rev. Res.}\ }\textbf {\bibinfo {volume} {2}},\
  \bibinfo {pages} {043416} (\bibinfo {year} {2020})}\BibitemShut {NoStop}%
\bibitem [{\citenamefont {Yepez}(2011)}]{Yepez2011}%
  \BibitemOpen
  \bibfield  {author} {\bibinfo {author} {\bibfnamefont {J.}~\bibnamefont
  {Yepez}},\ }\href@noop {} {\bibinfo {title} {Einstein's vierbein field theory
  of curved space}} (\bibinfo {year} {2011}),\ \Eprint
  {https://arxiv.org/abs/arXiv:1106.2037} {arXiv:1106.2037} \BibitemShut
  {NoStop}%
\bibitem [{\citenamefont {Witten}(2016)}]{Witten2016}%
  \BibitemOpen
  \bibfield  {author} {\bibinfo {author} {\bibfnamefont {E.}~\bibnamefont
  {Witten}},\ }\bibfield  {title} {\bibinfo {title} {Three lectures on
  topological phases of matter},\ }\href
  {https://doi.org/10.1393/ncr/i2016-10125-3} {\bibfield  {journal} {\bibinfo
  {journal} {La Rivista del Nuovo Cimento}\ }\textbf {\bibinfo {volume} {39}},\
  \bibinfo {pages} {313–370} (\bibinfo {year} {2016})}\BibitemShut {NoStop}%
\bibitem [{\citenamefont {Al-Marzoog}\ \emph {et~al.}(2024)\citenamefont
  {Al-Marzoog}, \citenamefont {Rezaei}, \citenamefont {Noorinejad},
  \citenamefont {Amini}, \citenamefont {Ghanbari-Adivi},\ and\ \citenamefont
  {Jafari}}]{AlMarzoog}%
  \BibitemOpen
  \bibfield  {author} {\bibinfo {author} {\bibfnamefont {R.}~\bibnamefont
  {Al-Marzoog}}, \bibinfo {author} {\bibfnamefont {A.}~\bibnamefont {Rezaei}},
  \bibinfo {author} {\bibfnamefont {Z.}~\bibnamefont {Noorinejad}}, \bibinfo
  {author} {\bibfnamefont {M.}~\bibnamefont {Amini}}, \bibinfo {author}
  {\bibfnamefont {E.}~\bibnamefont {Ghanbari-Adivi}},\ and\ \bibinfo {author}
  {\bibfnamefont {S.~A.}\ \bibnamefont {Jafari}},\ }\bibfield  {title}
  {\bibinfo {title} {Tilted dirac cones in two-dimensional materials: Impact on
  electron transmission and pseudospin dynamics},\ }\href
  {https://doi.org/10.1103/PhysRevB.110.165427} {\bibfield  {journal} {\bibinfo
   {journal} {Phys. Rev. B}\ }\textbf {\bibinfo {volume} {110}},\ \bibinfo
  {pages} {165427} (\bibinfo {year} {2024})}\BibitemShut {NoStop}%
\bibitem [{\citenamefont {Howard}(1992)}]{GeorgiWaves}%
  \BibitemOpen
  \bibfield  {author} {\bibinfo {author} {\bibfnamefont {G.}~\bibnamefont
  {Howard}},\ }\href@noop {} {\emph {\bibinfo {title} {The Physics of Waves}}}\
  (\bibinfo  {publisher} {Benjamin Cummings},\ \bibinfo {year}
  {1992})\BibitemShut {NoStop}%
\bibitem [{\citenamefont {Gong}\ \emph {et~al.}(2010)\citenamefont {Gong},
  \citenamefont {Zhao}, \citenamefont {Rao}, \citenamefont {Wu},\ and\
  \citenamefont {Guo}}]{Gong2010}%
  \BibitemOpen
  \bibfield  {author} {\bibinfo {author} {\bibfnamefont {Y.}~\bibnamefont
  {Gong}}, \bibinfo {author} {\bibfnamefont {T.}~\bibnamefont {Zhao}}, \bibinfo
  {author} {\bibfnamefont {Y.-J.}\ \bibnamefont {Rao}}, \bibinfo {author}
  {\bibfnamefont {Y.}~\bibnamefont {Wu}},\ and\ \bibinfo {author}
  {\bibfnamefont {Y.}~\bibnamefont {Guo}},\ }\bibfield  {title} {\bibinfo
  {title} {A ray-transfer-matrix model for hybrid fiber fabry-perot sensor
  based on graded-index multimode fiber},\ }\href
  {https://doi.org/10.1364/oe.18.015844} {\bibfield  {journal} {\bibinfo
  {journal} {Optics Express}\ }\textbf {\bibinfo {volume} {18}},\ \bibinfo
  {pages} {15844} (\bibinfo {year} {2010})}\BibitemShut {NoStop}%
\bibitem [{\citenamefont {Gong}\ \emph {et~al.}(2014)\citenamefont {Gong},
  \citenamefont {Huang}, \citenamefont {Liu}, \citenamefont {Wu}, \citenamefont
  {Rao}, \citenamefont {Peng}, \citenamefont {Lang},\ and\ \citenamefont
  {Zhang}}]{Gong2014}%
  \BibitemOpen
  \bibfield  {author} {\bibinfo {author} {\bibfnamefont {Y.}~\bibnamefont
  {Gong}}, \bibinfo {author} {\bibfnamefont {W.}~\bibnamefont {Huang}},
  \bibinfo {author} {\bibfnamefont {Q.-F.}\ \bibnamefont {Liu}}, \bibinfo
  {author} {\bibfnamefont {Y.}~\bibnamefont {Wu}}, \bibinfo {author}
  {\bibfnamefont {Y.}~\bibnamefont {Rao}}, \bibinfo {author} {\bibfnamefont
  {G.-D.}\ \bibnamefont {Peng}}, \bibinfo {author} {\bibfnamefont
  {J.}~\bibnamefont {Lang}},\ and\ \bibinfo {author} {\bibfnamefont
  {K.}~\bibnamefont {Zhang}},\ }\bibfield  {title} {\bibinfo {title}
  {Graded-index optical fiber tweezers with long manipulation length},\ }\href
  {https://doi.org/10.1364/oe.22.025267} {\bibfield  {journal} {\bibinfo
  {journal} {Optics Express}\ }\textbf {\bibinfo {volume} {22}},\ \bibinfo
  {pages} {25267} (\bibinfo {year} {2014})}\BibitemShut {NoStop}%
\end{thebibliography}%



\end{document}